\documentclass[%
 reprint,
 amsmath,amssymb,
 aps,
 pra
]{revtex4-1}

\usepackage{times}
\usepackage{txfonts}
\usepackage{graphicx}
\usepackage{dcolumn}
\usepackage{bm}
\usepackage{color}
\usepackage{hyperref}
\usepackage[T1]{fontenc}
\usepackage[utf8]{inputenc}
\hypersetup{
	unicode=true,          
	pdftitle={Interaction control and molecule production in the RbHg system},    
	pdfauthor={Mateusz Borkowski et al.},     
	colorlinks=true,       
	linkcolor=blue,          
	citecolor=blue,        
	filecolor=blue,      
	urlcolor=blue,           
}

\usepackage{natbib} 
\usepackage{color}

\begin{document}

\preprint{APS/123-QED}

\title{Optical Feshbach resonances and ground state molecule production in the RbHg system}

\author{Mateusz Borkowski}
\email{mateusz@fizyka.umk.pl}
\author{Rodolfo Muñoz Rodriguez}
\author{Maciej B. Kosicki}
\author{Roman Ciuryło}
\author{Piotr S. Żuchowski}%
\email{pzuch@fizyka.umk.pl}
\affiliation{Institute of Physics, Faculty of Physics, Astronomy and Informatics, Nicolaus Copernicus University, Grudziadzka 5, 87-100 Torun, Poland}

\date{\today}
            
\begin{abstract}
We present the prospects for photoassociation, optical control of interspecies scattering lengths and finally, the production of ultracold absolute ground state molecules in the Rb+Hg system. We use the ,,gold standard'' \emph{ab initio} methods for the calculations of ground (CCSD(T)) and excited state (EOM-CCSD) potential curves. The RbHg system, thanks to the wide range of stable Hg bosonic isotopes, offers possibilities for mass-tuning of ground state interactions. The optical lengths describing the strengths of optical Feshbach resonances near the Rb transitions are favorable even at large laser detunings. Ground state RbHg molecules can be produced with efficiencies ranging from about 20\% for deeply bound to at least 50\% for weakly bound states close to the dissociation limit. Finally, electronic transitions with favorable Franck-Condon factors can be found for the purposes of a STIRAP transfer of the weakly bound RbHg molecules to the absolute ground state using commercially available lasers.
\end{abstract}


\maketitle

\section{Introduction \label{sec:Introduction}}

The increasing popularity, in recent years, of the research on ultracold molecules is fueled by their possible intriguing applications in diverse areas of physics. Pioneering experiments with molecules packed in 3D optical lattices reported recently~\cite{Moses2015} raise hope for using ultracold molecules trapped in periodic potentials as quantum simulations of condensed matter physics Hamiltonians~\cite{Baranov2012,Buechler2007,Micheli2006}. Ultracold molecules offer the possibility to the study of chemistry under yet unexplored conditions: at extremely low energies and in controllable quantum states, and where the chemical reaction could be manipulated with external magnetic, or electric fields. Recent experimental realizations include reactive collisions of KRb molecules~\cite{Ospelkaus2010}, or the photodissociation of Sr$_2$ molecule~\cite{McDonald2016}. Finally, ultracold molecules offer new avenues and improved precision in experimental searches of ,,new physics''. Cold molecules are already being used for the determination of experimental constraints on the electric dipole moment of the electron~\cite{Hudson2011} and the time drift of the fine-structure constant~\cite{Truppe2013,Hudson2006}. Their production at microkelvin temperatures might improve these experiments.
Ultracold samples of molecules can be formed from ultracold atoms by magnetoassociation \cite{Chin2010} followed by an adiabatic transfer by two lasers (STIRAP) \cite{Bergmann1998} from the weakly bound state to deeply bound electronic ground-states~\cite{Sage2005,Ni2008,Danzl2010}. At present, molecules obtained by this procedure include KRb~\cite{Ni2008}, NaK~\cite{Wu2014}, RbCs~\cite{Molony2014,Takekoshi2014} and NaRb~\cite{Guo2016}.  

Apart from progress in the formation of ultracold alkali-metal dimers, there are ongoing efforts on formation of new types of molecules, in particular heteronuclear, open-shell molecules. Such molecules should provide more opportunities for external field control as they possess not only electric, but also magnetic dipole moments and offer more possibilities of control of their state and properties. Also, paramagnetic polar molecules were proposed by Micheli {\em et al.}~\cite{Micheli2006}  for the creation of topologically ordered states with possible use in quantum computing. Recently, interest in these systems was also boosted by reporting the mechanisms which might allow to form weakly bound Feshbach molecules via  magnetoassociation~\cite{Zuchowski2010,Brue2012, Tomza2014}. It is therefore no surprise that several research groups are pursuing experiments on ultracold  mixtures of alkali-metal and closed-shell atoms. The earliest experiments were conducted with the Rb+Yb system, for which the first working magneto-optical trap was reported~\cite{Tassy2010,Baumer2011} and  one-color~\cite{Nemitz2009} and two-color photoassociation spectroscopy~\cite{Munchow2011,Borkowski2013} experiments were performed. The latter which made it possible to acquire information about the short-range potential energy curves (PECs) for this system. The Rb+Sr system is currently under systematic investigation, in particular the production of a binary mixture of Bose-Einstein condensates was reported~\cite{Pasquiou2013}. For the Cs+Yb system a dual magneto-optical trap was created~\cite{Kemp2016} and, more recently, the interspecies thermalization properties of the mixture in an optical trap~\cite{Guttridge2017} were investigated. The cotrapping of Li+Yb mixtures was studied by Hara {\em et al.}~\cite{Hara2011} and Hansen {\em et al.}~\cite{Hansen2013}. Interestingly, magnetic Feshbach resonances in collisions of electronically excited Yb($^3P_2$) and Li were reported. The Li+Yb system is currently being studied using photoassociation spectroscopy close to the lithium $^2$S$\to$$^2$P transition~\cite{Roy2016}.

Our work is motivated by the experimental progress in the trapping of Rb and Hg atoms in a dual magneto-optical trap (MOT) which was made recently in our group~\cite{Witkowski2017}: approximately $10^6$ Rb atoms were trapped simultaneously with about $10^5$ Hg atoms. Among atoms that can be laser cooled, Hg has particularly interesting properties. It is an appealing building block for a new generation of optical clocks, due to very low black-body radiation pumping-related losses. Because of its large mass, it is a good candidate for parity-violation studies. The effort to simultaneously cool Rb and Hg atoms despite the experimental challenges -- for example, the deep UV 254~nm wavelength used for Doppler cooling of Hg -- is driven by the possibility of obtaining the grand prize of ultracold RbHg molecules. Dimers containing the Hg atom were proposed by Meyer and Bohn as appropriate candidates for the search for the electric dipole moment of the electron~\cite{Meyer2009}, which is due to the fact that Hg is among the heaviest atoms which can be laser cooled.

Dimers containing the Hg atom have been studied in past. Bound-bound transitions in the Hg$_2$ molecule were investigated with fluorescence spectroscopy in supersonic beams and was subject to experiment-theory comparisons~\cite{Krosnicki2015}. Hg$_2$ was the subject of femtosecond photoassociation spectroscopy~\cite{Marvet1995}. The ground state mercury interaction was also thoroughly studied in the context of bulk properties~\cite{Gaston2006}. Because of its importance high-quality effective core potentials (ECP) for Hg atom were tailored by Dolg and Stoll \cite{Figgen2005} and a family of correlation--consistent basis sets were tailored by Peterson and Puzzarini \cite{Peterson:05}. The agreement between spectroscopic data and quantum chemistry calculations was very good.
Very little, however, is known about the interaction of alkali-metal atoms with mercury. The best studied system to date is Li+Hg, for which the results of bound-bound and bound-free fluorescence spectroscopy were corroborated by high-quality quantum chemical calculations of the lowest excited states~\cite{Gruber1994,Gruber1996}. High-quality relativistic studies of the CsHg system were performed by~Polly~\emph{et~al.}~\cite{Polly1998}.

\begin{figure}[t]
	\includegraphics[width=0.5\textwidth]{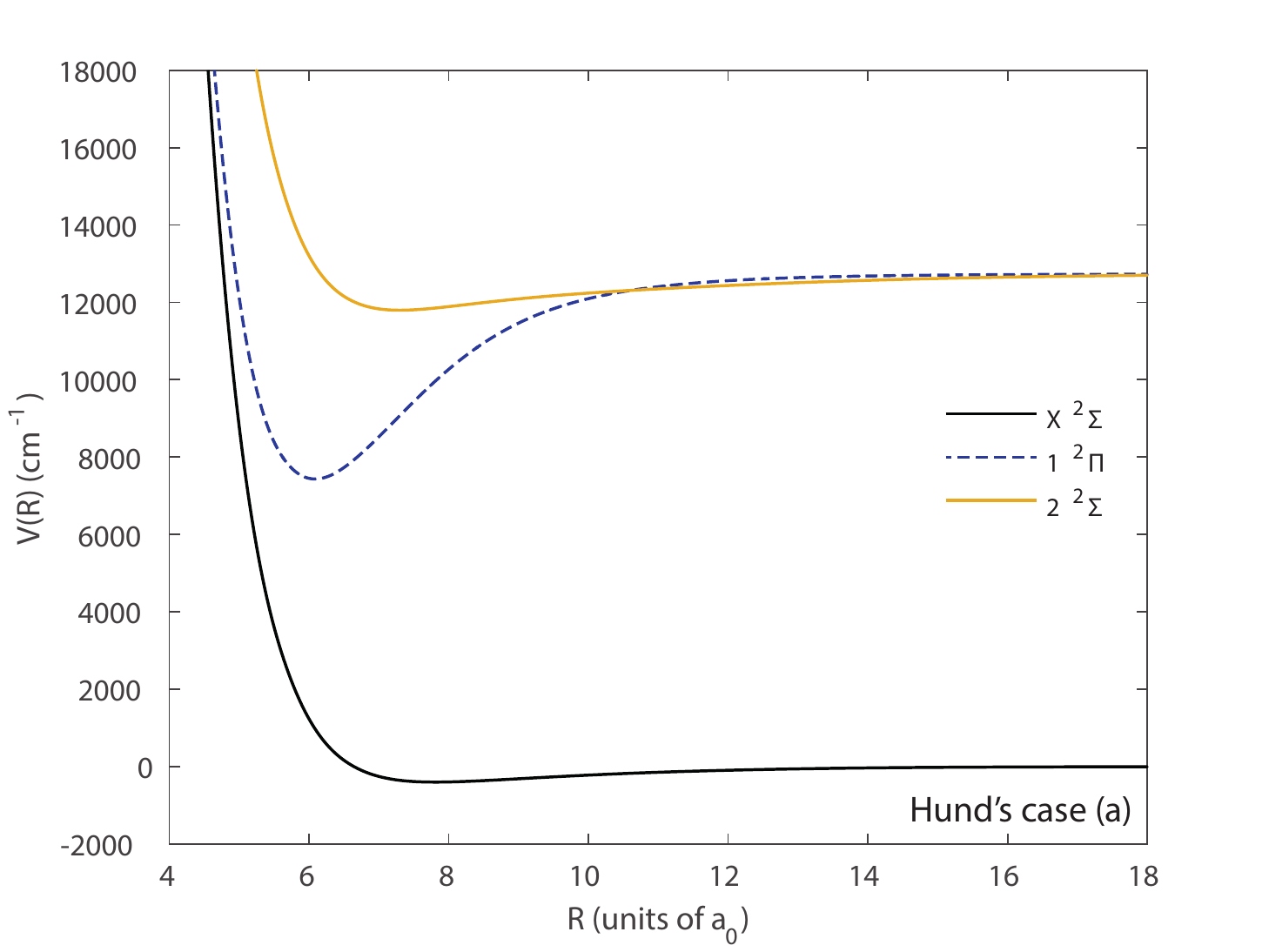}
	\caption{(Color online) Potential energy curves near the $^2$P+$^1$S and $^2$S+$^1$S asymptotes of RbHg in Hund's case (a) representation calculated using state-of-the-art~\emph{ab initio} methods, see Sec.~\ref{sec:ab_initio} for details.}
	\label{fig:PEC_HundA}
\end{figure}

\begin{figure}[t]
	\includegraphics[width=0.505\textwidth]{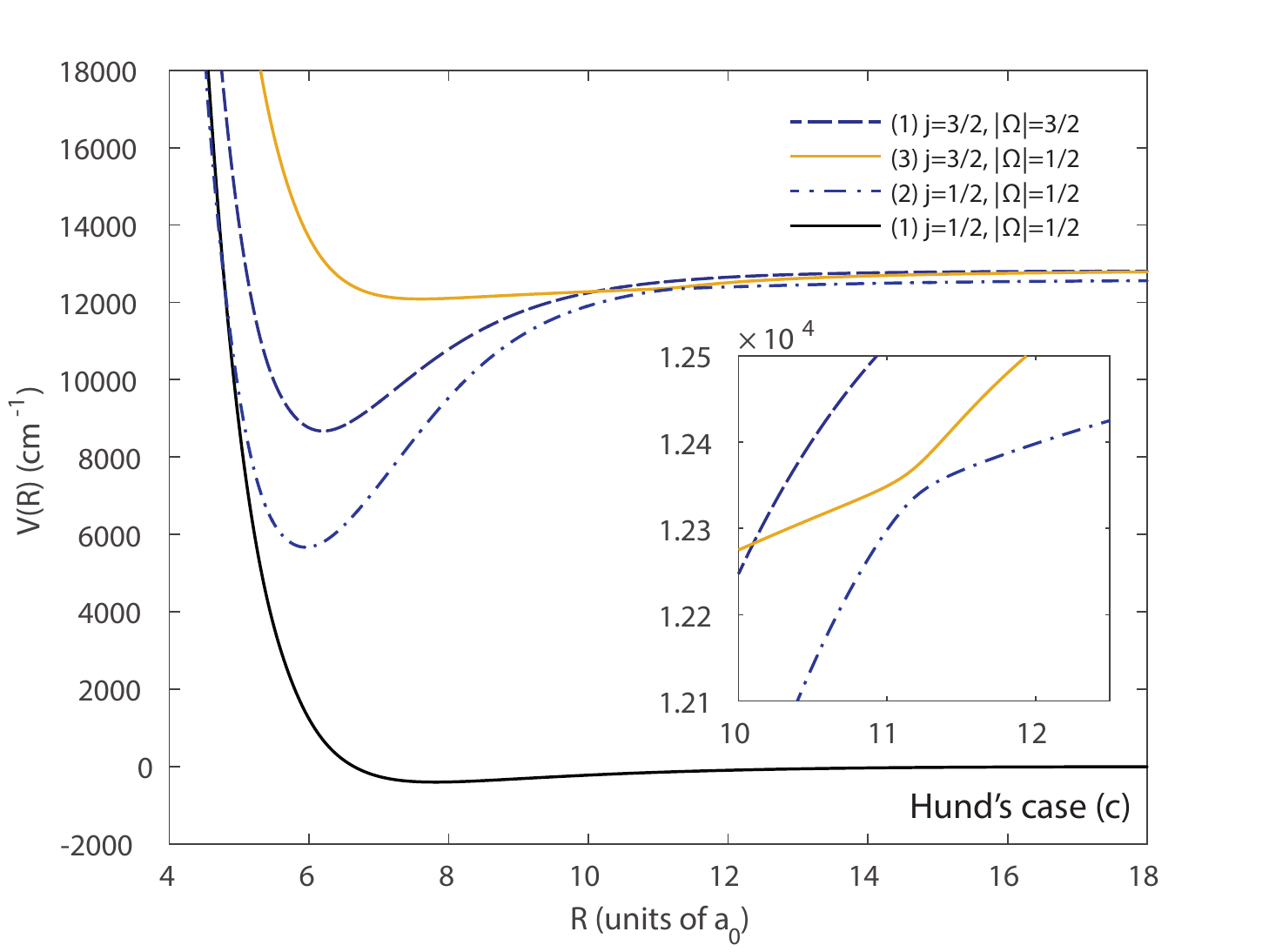}
	\caption{(Color online) Potential energy curves near the $^2$P+$^1$S and $^2$S+$^1$S asymptotes of RbHg in the L-S coupled Hund's case (c) representations calculated using Eq.~(\ref{eq:HhundC}) from the nonrelativistic \emph{ab initio} potentials shown in Fig.~\ref{fig:PEC_HundA} and the spin-orbit matrix elements shown in Fig.~\ref{fig:matrixelements}a.
		\label{fig:PEC_HundC}}
\end{figure}

Photoassociation experiments using the Rb D1 and D2 transitions in RbHg are underway. This work addresses the possibilities of such photoassociation, the main features of the photoassociation resonances, and the possibility of manipulating the collisional properties of the Rb+Hg mixture in ultracold regime by optical Feshbach resonance \cite{Fedichev1996, Bohn1997, Ciurylo2005, Kokoouline2001}.
We also investigate the products of spontaneous emission during photoassociation for the purposes of electronic ground-state molecule formation via stimulated Raman adiabatic passage (STIRAP) \cite{Bergmann1998, Ni2008}. The paper is organized as follows. In the next section we describe in detail the methodology used in the calculations of the potential energy curves, for ground and excited states. In section~\ref{sec:couplings} we discuss the interactions of Rb and Hg near the first two asymptotes of the excited Rb atom and provide the transition from the non-relativistic Hund's case (a) to the spin-orbit coupled Hund's case (c) framework, as well as Hund's case (e) which includes the coupling with the rotational angular momentum. In Sec.~\ref{sec:scatlen} we provide an analysis of the scattering properties of the RbHg system (for all possible isotopic combinations) and, in Sec.~\ref{sec:l_opt}, explore the possibilities of tuning the scattering length in this system by optical Feshbach resonances. Section \ref{sec:molecule_production} is focused on studies of the bound-bound transitions and Franck-Condon factors in RbHg between the ground-state and lowest excited states correlating with $^2P$ state of the Rb. In particular, we discuss the possibility of molecule formation by spontaneous emission from the photoassociatively formed excited molecular state to the ground state and opportunities for STIRAP transfers to the vibrational ground state. Section~\ref{sec:summary} concludes our paper.

\section{Ab-initio Interaction potentials \label{sec:ab_initio}}

To study the Born-Oppenheimer  interaction energy  in the ground state we have performed high-level \emph{ab initio} calculations using the spin-restricted open-shell coupled cluster method with single, double, and noniterative triple excitations CCSD(T)~\cite{Knowles1993} implemented in the \textsc{molpro} package~\cite{MOLPRO_brief:2012}.
The excited-state potential energy curves are obtained using the EOM-CCSD method~\cite{Stanton1993} for the calculations of excitation energies which were subsequently added to the ground state potentials. These \emph{ab initio} methods provide us with Hund's case (a) potentials.  In order to convert them to the relativistic, spin-orbit coupled Hund's case (c) picture we employ the spin-orbit coupling matrix elements calculated using the multireference configuration theory (MRCI). A similar methodology was used in the recent studies of the 
Ca$_2$~\cite{Bussery-Honvault2006} and Sr$_2$~\cite{Skomorowski2012}~systems and the interactions of alkali-metal atoms with strontium and calcium~\cite{Pototschnig2017}. 

We will first focus on the ground-state potential calculations. The CCSD(T) calculations were performed using the \textsc{molpro} package~\cite{MOLPRO_brief:2012}.  We applied the counter-poise correction (CP) proposed by Boys and Bernardi \cite{Boys:1970} to eliminate the basis set superposition error (BSSE) in the interaction energy.  Both the small-core relativistic energy-consistent pseudopotentials (ECP) as well as the full tailored valence basis sets optimized by Lim \textit{et al.}  \cite{Lim2005} were used, with additional \textit{h} functions and diffused functions added by us to better describe the Rb atom (see Ref.~\cite{Zuchowski2014} for details). 
The Hg atom was described by the augmented correlation consistent polarized valence quadruple-zeta quality basis sets (aug-cc-pVQZ-PP) optimized by Peterson and Puzzarini \cite{Peterson:05} with ECP of Figgen et al.~\cite{Figgen2005}.  To better account for the dispersion interactions we  added \textit{spdf} midbond functions held in the center of mass between both monomers.

The basis sets were tested for convergence by performing calculations  with triple-zeta basis sets for the Hg atom and the basis sets for Rb truncated at the $g$ functions (we denote both basis sets as TZ).  The complete basis set limit (CBS) of the ground-state potential was estimated by extrapolation from triple- and quadruple-zeta quality functions~\cite{Helgaker:2000}. For the scattering and bound-state calculations we used the quadruple-zeta basis potentials, whereas CBS gives us an indication of the possible error related to basis set incompleteness. Table~\ref{tab:PES} collects the potential depths for both basis sets and CBS. The $D_e$ parameter for the QZ quality basis set is 404 cm$^{-1}$ while the extrapolation to the basis set limit yields 412 cm$^{-1}$ which indicates a small uncertainty of our potential related to the basis truncation.

\begin{table}[b]
	\caption{Convergence of the interaction energy potential depth ($D_e$) and the equilibrium bond length ($R_e$) of RbHg molecule in the electronic ground-state. The energy unit is cm$^{-1}$ while the equilibrium distance is given in $a_0$.   \label{tab:PES}}
	\begin{ruledtabular}
		\begin{tabular}{lll}
			Basis set & $R_e$ & $D_e$    \\
			\hline
			TZ & 7.83 & 393 \\ 
			QZ & 7.82 & 404 \\
			CBS limit (estimate)& 7.80 & 412 \\
		\end{tabular}
	\end{ruledtabular}
\end{table}

\begin{table}[b]
	\begin{ruledtabular}
		\caption{Spectroscopic properties of Hund's case (a) and Hund's case (c) potential curves used in this analysis. Here the harmonic constant $\omega$ is defined as the energy difference between the bottom two vibrational states for the lowest rotational state. We also give the number of supported vibrational states $N$ for each of the potential curves.}
		\label{tab:PES2}
		\begin{tabular}{l r r r r r}
			State & $D_e$ (cm$^{-1}$) & $R_e$ ($a_0$) & $\omega$ (cm$^{-1}$) & $C_6$ ($E_h a_0^6$) & $N$ \\
			\hline
			2~$^2\Sigma$ & 940 & 7.31 & 34.1 & 2656 & 79 \\
			1~$^2\Pi$   & 5304 & 6.10 & 91.2 & 1440 & 117 \\
			X $^2\Sigma$ & 404 & 7.82 & 21.4 &  949.7 & 44 \\		
			\hline
			(1) j=3/2, $|\Omega|$=3/2 & 4143 & 6.19 & 85.6 & 1440 & 104 \\
			(3) j=3/2, $|\Omega|$=1/2 & 728   & 7.61 & 22.9 & 2251 & 73 \\
			(2) j=1/2, $|\Omega|$=1/2 & 6911 & 5.95 & 97.4 & 1845 & 135 \\
		\end{tabular}	
	\end{ruledtabular}
\end{table}

The basis set uncertainty is not the only one we deal with, and other uncertainties include the correlation energy beyond CCSD(T), as well as the relativistic effects which here are described only via the ECP. To assess the quality of  the  calculated CCSD(T) potential depth for the RbHg system we compared the well depth in the  CBS limit for the Hg$_2$ system, which was examined experimentally and by fully relativistic electronic structure methods. The CBS limit for the Hg$_2$ potential well obtained by us, $390.1$ cm${}^{-1})$, is very close to the dissociation energy $(383.4$ cm${}^{-1})$ found using the quadruple-zeta-quality basis sets (for both cases $R_e$ was found at 7.0 $a_0$). These values are in a very good agreement with the experimental data reported by Koperski $(379.5$ cm${}^{-1} $ and $6.8$ $a_0)$, \cite{Koperski:94, Koperski:97, Koperski:08}, and CC calculations including the full triply-excited configurations and the spin-orbit coupling corrections obtained by Schwerdtfeger $(392$ cm${}^{-1}$ and $6.95$ $a_0)$~\cite{Figgen:11}. Clearly, the interaction of mercury is reproduced extremely well with the ECP and basis set used in this paper, hence it should be trusted for RbHg calculations as well. We conservatively estimate the error on the potential depth to be on the order of 20~cm$^{-1}$.

The calculations of lowest excited states were performed as follows. Using the same ECPs as for the ground state we have performed EOM-CCSD calculations using the {\sc cfour}~\cite{CFOUR_brief} program. The basis sets used in the EOM-CCSD calculations were smaller: for the Hg atom we used the uncontracted aug-cc-pVTZ-PP \cite{Peterson:05} basis set and the Rb basis set restricted to $g$ functions  and the midbond  functions were not added.  The EOM-CCSD method calculates the energy difference between the ground state and pertinent excited states which are of our interest: the first $\Pi$ and the second $\Sigma$ states. To obtain the potential energy curves we added the calculated energy differences to the  ground state CCSD(T) potential. The asymptotic limit of the excitation energy agrees between the  $\Sigma$ and $\Pi$ (12797 cm$^{-1}$) states and both agree very well with the non-relativistic limit of the Rb atom $^2$P asymptote of 12737 cm$^{-1}$ which can be derived from the experimental values assuming Land\'e splittings. We also obtain a very good agreement between EOM-CCSD transition dipole moments and their experimental values (3.01 compared to the experimental 2.99 a.u.).

In Fig.~\ref{fig:PEC_HundA} we show Hund's case (a) curves which were used in dynamic calculations in present work, while Table~\ref{tab:PES2} gathers their spectroscopic properties. The depth of the ground-state curve is very low, given that RbHg represents the interaction of two metal atoms. It is about two times shallower than the potential wells in RbYb and nearly three times shallower than the comparable RbSr system~\cite{Zuchowski2014,Pototschnig2017,Borkowski2013}.
RbHg (similarly to RbYb and RbSr) is unbound at the Hartree-Fock level and it is only the dispersion energy that binds the molecule. 

The depths of the potential wells for RbSr, RbYb, RbHg systems show decreasing strengths of the dispersion interaction, which can be rationalized by comparing the polarizabilities of these atoms (195 for Sr, 143 for Yb, and 35 a.u. for Hg). Interestingly, Polly et al.~\cite{Polly1998} have obtained a very shallow ground-state potential for the CsHg molecule (160 cm$^{-1}$) but such disagreement might be explained by the fact that no triply excited configurations were included by these authors.

The excited states of the RbHg molecule are quite peculiar: the  1~$^2\Pi$ state is rather deep at 5304~cm$^{-1}$ but the 2~$^2\Sigma^+$ state (940~cm$^{-1}$) is, similarly to the ground state, very shallow compared to RbSr, RbYb or the recently reported RbCa systems~\cite{Zuchowski2014,Sorensen2009,Pototschnig2017}. Nonetheless, the long-range interaction for the $2$~$^2\Sigma^+$ state is stronger than in the case of  1~$^2\Pi$, hence these states cross at about 11 $a_0$ which produces an avoided crossing in the Hund's case (c) picture (see inset of Fig.~\ref{fig:PEC_HundC}).

To solve the radial Schr\"odinger equation for the ultracold regime and discuss the near-threshold bound states it is essential to use analytic van der Waals potentials at large internuclear separations. To this end we smoothly connected our potentials to analytical expansions with $C_6$ coefficients obtained from perturbation theory.  To obtain the $C_6$ coefficients for the ground state we used a Casimir-Polder type integral expressed in terms of the atomic dynamic dipole polarizabilities at imaginary frequencies (see Ref.~\cite{Zuchowski2014,Borkowski2013})
\begin{equation}
C_{6} = {\frac{3}{\pi}} \int_{0}^{\infty} \alpha_{A}(i\omega) \alpha_{B}(i\omega) d\omega.
\end{equation} 
The values of atomic dynamic electric dipole polarizabilities for the Rb atom were taken from Ref.~\cite{Derevianko:2010}. For the polarizabilities of the Hg atom, we have employed the time-independent coupled-cluster polarization propagator method (TI-CC)~\cite{ Moszynski:2005,Korona:06}. Since the dynamic polarizabilities for excited Rb atom in the $^2$P state were unavailable, we used the $C_6$ values for BeRb system~\cite{Jiang2013} (2988 and 1620 a.u.) scaled by the ratio of static polarizabilities of the Hg and Be atoms equal to  0.89.

The calculation of Hund's case (c) potential curves shown in Fig.~\ref{fig:PEC_HundC} requires the use of spin-orbit coupling matrix elements between the lowest excited $\Sigma$ and $\Pi$ states. The spin-orbit matrix elements were obtained using the MRCI method restricted to single and double excitations in the Breit-Pauli approximation as implemented in the \textsc{molpro} package~\cite{MOLPRO_brief:2012}. A large active space in the MRCI calculations included the \textit{s}, \textit{p}, and \textit{d} orbitals from the external electronic subshells. The basis sets for the Rb atom and the aug-cc-pVQZ-PP basis set for Hg atom were restricted to $d$ functions. Similarly to the EOM-CCSD calculations, the midbond functions were not added. The asymptotic spin-orbit matrix element matched the experimental Rb spin-orbit constant of 79.2~cm$^{-1}$ to about 5\%. The spin-orbit matrix elements for the 2~$^2\Sigma^+$ and 1~$^2\Pi$ states are shown in Fig.~\ref{fig:matrixelements}a. 

Finally, we have also calculated the dipole moment for the ground-state using the finite-field method~\cite{Pople1968,Zuchowski2013}. The RbHg molecule in its rovibrational ground-state has a comparatively small dipole moment of 0.056 $e a_0$.

\section{Coupling schemes \label{sec:couplings}}

\begin{figure}[t]
	\includegraphics[width=0.465\textwidth]{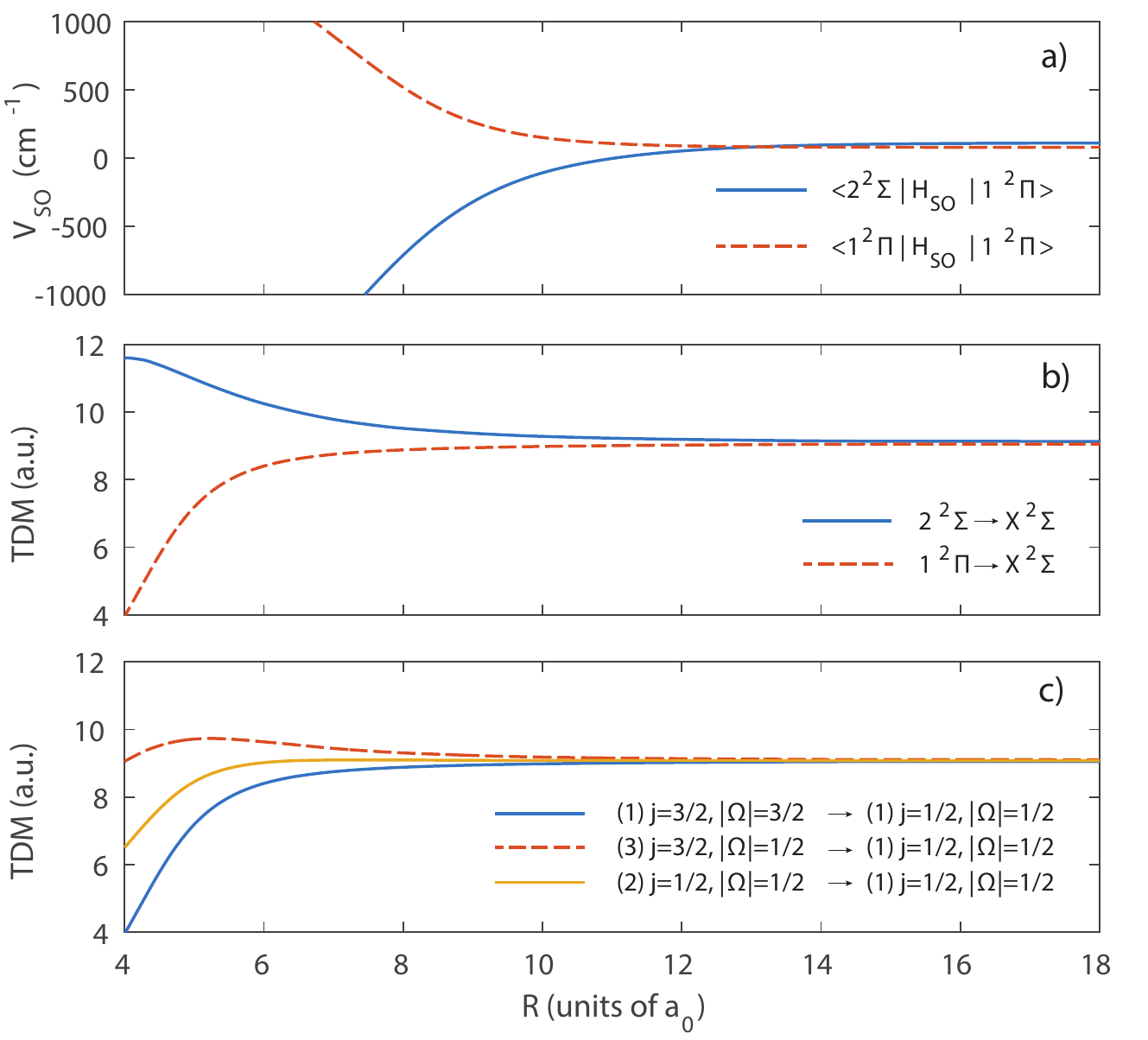}
	\caption{(Color online) a) Relevant spin-orbit matrix elements for Hund's case (a) potential curves correlating to the $^2$P-$^1$S asymptote of RbHg; b) and c) transition dipole moments from excited states to the ground state in Hund's case (a) and (c) representations, respectively.
		\label{fig:matrixelements}}
\end{figure}

Compared to RbYb or RbSr, the RbHg system has an entirely different
structure of the excited state thresholds. The Hg atom has a very high
excitation energy: the first strong optical transition to an excited state of Hg, the 254 nm intercombination $^1$S$_0$$\rightarrow$$^3$P$_1$ line used in laser cooling of Hg atoms, leads to an electronic state significantly above the ionization threshold of the Rb atom. Hence, electronic states of RbHg, where the Hg atom is excited, are coupled to a continuum of ionized Rb states.
Penning ionization of Rb atoms colliding with excited Hg atoms is to be expected. Theoretical description of such states is very challenging. Also, photoassociation and optical manipulation of such systems using the Hg optical transitions will be jeopardized by strong autoionization losses. Hence
our main focus in this paper is to consider photoassociation near transitions to the lowest $^2P$ Rb excited states.

The electronic configuration of the Rb atom in the first excited state, $5s5p$, is split by spin-orbit coupling
into $j=\frac{3}{2}$ (D2-line) and $j=\frac{1}{2}$ (D1-line) states. The
interaction with the Hg atom lifts the degeneracy of the $j=\frac{3}{2}$ state, creating two pairs of degenerate states 
$\Omega = \pm \frac{1}{2}, \pm \frac{3}{2}$, while the $j=\frac{1}{2}$ state produces one doubly degenerate state $\Omega=\pm \frac{1}{2}$,
where the quantum number $\Omega$ is the projection of the total atomic angular momentum onto the molecular axis. 
In the non-relativistic framework corresponding to the Hund's case (a), the molecular states are labeled by the molecular axis projections $\Sigma$ and $\Lambda$ of the spin and orbital angular momenta, respectively. 
Our {\em ab initio} calculations provide potential energy curves corresponding to a Hund's case (a) picture where the $^2$P state splits into one $^2\Sigma^+$ state and a doubly degenerate $^2\Pi$ state.

Similar to Ref. \cite{Ciurylo2004} we start from the quantum theory of slow-atom collisions \cite{Mies1973,Gao1996}, and describe the colliding system using the Hamiltonian 
\begin{equation}
H=T+H_{\rm A}+V_{\rm int}+V_{\rm rot}.
\end{equation}$T=(\hbar^2/2\mu)(d^2/dR^2)$ is the
kinetic energy operator for the relative radial motion, $H_{\rm A}$ is the
atomic Hamiltonian operator representing the internal atomic degrees of
freedom, $V_{\rm int}$ is the interaction operator described by
nonrelativistic molecular Born-Oppenheimer potentials, and $V_{\rm rot}$
is the rotational energy operator. While  discussing the interactions
correlating with the asymptotes Rb($^2S_\frac{1}{2}$)+Hg($^1S_0$),
Rb($^2P_\frac{1}{2}$)+Hg($^1S_0$), and
Rb($^2P_\frac{3}{2}$)+Hg($^1S_0$), we are using the $|jlJM\rangle$
basis corresponding to Hund's (e) case: $\vec{j}$ is the total
electronic angular momentum, $\vec{l}$ is the rotational (end-over-end) angular
momentum, and $\vec{J}=\vec{j}+\vec{l}$ is the total angular
momentum. The projection of $\vec{J}$ on a space-fixed $z$ axis is
$M$, however one can neglect the possible dependence on $M$ in the absence of 
external fields. Hence, possible channels will be labeled 
as $|jlJ\rangle$.

The adiabatic Hund's case (c) potentials are obtained by the diagonalization of the Born-Oppenheimer potentials and the spin-orbit coupling operator $H_{\rm SO}$ at a given distance $R$. The Hund's case (c) potential for $|\Omega| = 3/2$
\begin{equation}
	V\left((1)\,j=3/2,\,|\Omega|=3/2\right) = V(1\, ^2\Pi) + \left<1\, ^2\Pi\left| H_{\rm SO} \right|1\, ^2\Pi \right> \,.
\end{equation}
The other two states 
for $\Omega=\pm\frac{1}{2}$ can be obtained by diagonalizing the following matrix~\cite{Zuchowski2014}:
\begin{equation}
H\left(|\Omega|=\frac{1}{2}\right) = \left(
\begin{array}{cc}
V(2\,^2\Sigma) &  \left<2\, ^2\Sigma\left| H_{\rm SO} \right|1\, ^2\Pi \right> \\
\left<2\, ^2\Sigma\left| H_{\rm SO} \right|1\, ^2\Pi \right>  &  V(1^2\Pi)  - \left<1\, ^2\Pi\left| H_{\rm SO} \right|1\, ^2\Pi \right>
\end{array}\right). \label{eq:HhundC}
\end{equation}
Fig.~\ref{fig:PEC_HundC} shows the spin-orbit coupled potential energy curves, while Table~\ref{tab:PES2} lists their essential molecular properties. The spin-orbit matrix elements $\left<2\, ^2\Sigma\left| H_{\rm SO} \right|1\, ^2\Pi \right>$ and $\left<1\, ^2\Pi\left| H_{\rm SO} \right|1\, ^2\Pi \right>$ are shown in Fig.~\ref{fig:matrixelements}a; their respective asymptotic limits are $\sqrt{2} A_{\rm Rb}$ and $A_{\rm Rb}$, where the Rb spin-orbit constant $A_{\rm Rb}=79.2$~cm$^{-1}$ is equal to one third of the energy difference between the atomic $^2P_{\frac{1}{2}}$ and $^2P_{\frac{3}{2}}$ states. In the asymptotic limit, where the spin-orbit coupling dominates over the atomic interactions, the potentials are strongly mixed:
in excited electronic states near the Rb($^2P_\frac{3}{2}$)+Hg($^1S_0$) asymptote,
\begin{eqnarray}
V_e\left({\rm (1)\, j=3/2,|\Omega|=3/2}\right)&=&V_e(1~^2\Pi), \, {\rm and} \\
V_e\left({\rm (3)\, j=3/2,|\Omega|=1/2}\right)&=&\frac{2}{3}V_e(2~^2\Sigma)+\frac{1}{3}V_e(1~^2\Pi),
\end{eqnarray}
while near the Rb($^2P_\frac{1}{2}$)+Hg($^1S_0$) asymptote,
\begin{equation}
V_e\left({\rm (2)\, j=1/2,|\Omega|=1/2}\right)=\frac{1}{3}V_e(2~^2\Sigma)+\frac{2}{3}V_e(1~^2\Pi).
\end{equation}
This mixing determines the long range $C_6$ coefficients (see Table \ref{tab:PES2}) for the Hund's case potentials used in our analysis. At equilibrium distances, however, the shapes and depths of the Hund's case (c) potentials are mostly determined by the Hund's case (a) potentials. The (1)~j=3/2,~$|\Omega|$=3/2 and (2)~j=1/2,~$|\Omega|$=1/2 potentials inherit the shape of the 1~$^2\Pi$ curve while (3)~j=3/2,~$|\Omega|$=1/2 approximates the shallow 2~$^2\Sigma$ potential as seen in Figure~\ref{fig:PEC_HundC}.

The rotational energy operator $V_{\rm rot} = B(R)l(l+1)$ is diagonal in the Hund's case (e) but not in the Hund's case (c) representations which leads to possible rotational (Coriolis) couplings between Hund's case (c) channels. On the other hand, a Hund's case (c) state may mix many rotational states.
Due to selection rules, photoassociation in $s$-wave collisions can occur only to states including channels $|j~l~J\rangle$ with $l=0$. Such mixing will impact the photoassociation rate through an appropriate H\"onl-London factor $f_{\rm rot}$. 

States correlating to the Rb($^2S_{\frac{1}{2}}$)+Hg($^1S_0$) and
Rb($^2P_\frac{1}{2}$)+Hg($^1S_0$) asymptotes can be described by single channels $|jlJ\rangle$ in which $j=1/2$ and $l=J\pm1/2$. In ultracold collisions the $s$-wave channel $|j=1/2, l=0, J=1/2\rangle$
with rotational energy equal to zero plays the crucial role and higher
partial waves can be neglected. In this case $V_{\rm int}+V_{\rm rot}$ is simply
$V_g((1)~1/2_{1/2})$. In the excited Rb($^2P_\frac{1}{2}$)+Hg($^1S_0$) state photoassociation will be possible only to the bound states supported by the channel $|j=1/2, l=0, J=1/2\rangle$ where $V_{\rm int}+V_{\rm rot}$ is equal to $V_e((2)~j=1/2, |\Omega|={1/2})$. No inter-channel rotational couplings are present here.

The situation is more complex near the Rb($^2P_\frac{3}{2}$)+Hg($^1S_0$) asymptote where the bound states need to be represented by two channels. It can be shown that for $s$-wave collisions in this asymptote, photoassociation
is possible only to bound states described by a pair of the
following channels: $|j=3/2, l=0, |\Omega|=3/2\rangle$ and
$|j=3/2,l=2,|\Omega|=3/2\rangle$. Through a calculation similar to that of  Ref.~\cite{Borkowski2009} one can show that near
the dissociation limit two sets of Hund's case (c) bound states can be found.
For interaction energies much larger than the rotational energy, these can be described as single channels with effective $V_{\rm int}+V_{\rm rot}$ given by
\begin{eqnarray}
V_e((3)~j=3/2, |\Omega|=1/2)+3B &\longrightarrow& \left(
\begin{array}{c}
+\sqrt{1/2}\vspace{3mm}\\
-\sqrt{1/2}
\end{array}
\right)
\begin{array}{c}
l=0\vspace{3mm}\\
l=2
\end{array}
\label{eq:Pot3212} 
\\
{\rm and\vspace{3cm}}\nonumber
\\
V_e((1)~j=3/2, |\Omega|=3/2)+3B &\longrightarrow& \left(
\begin{array}{c}
+\sqrt{1/2}\vspace{3mm}\\
+\sqrt{1/2}
\end{array}
\right)
\begin{array}{c}
l=0\vspace{3mm}\\
l=2
\end{array}.
\label{eq:Pot3232}
\end{eqnarray}
The centrifugal term $B(R)=\hbar^{2}/(2\mu R^{2})$. The eigenvectors to the right denote the rotational composition of these Hund's case (c) states, which turn out to be 1:1 mixtures of the $s$ and $d$-waves.

In our work paper we neglect the hyperfine structure of the Rb atoms. The strength of coupling between nuclear and electronic spins is orders of magnitude smaller than the coupling between electronic spin and the orbital angular momentum. While the hyperfine interaction does have a minor impact on the shape of the potential curves~\cite{Zuchowski2010}, we expect it to be even smaller than that in RbYb \cite{Borkowski2013}. The qualitative results reported here will therefore remain valid regardless of the nuclear spin as long as no magnetic fields are used in experiment.

\section{s-wave scattering lengths \label{sec:scatlen}}

\begin{figure}[t]
	\includegraphics[clip, width=0.47\textwidth]{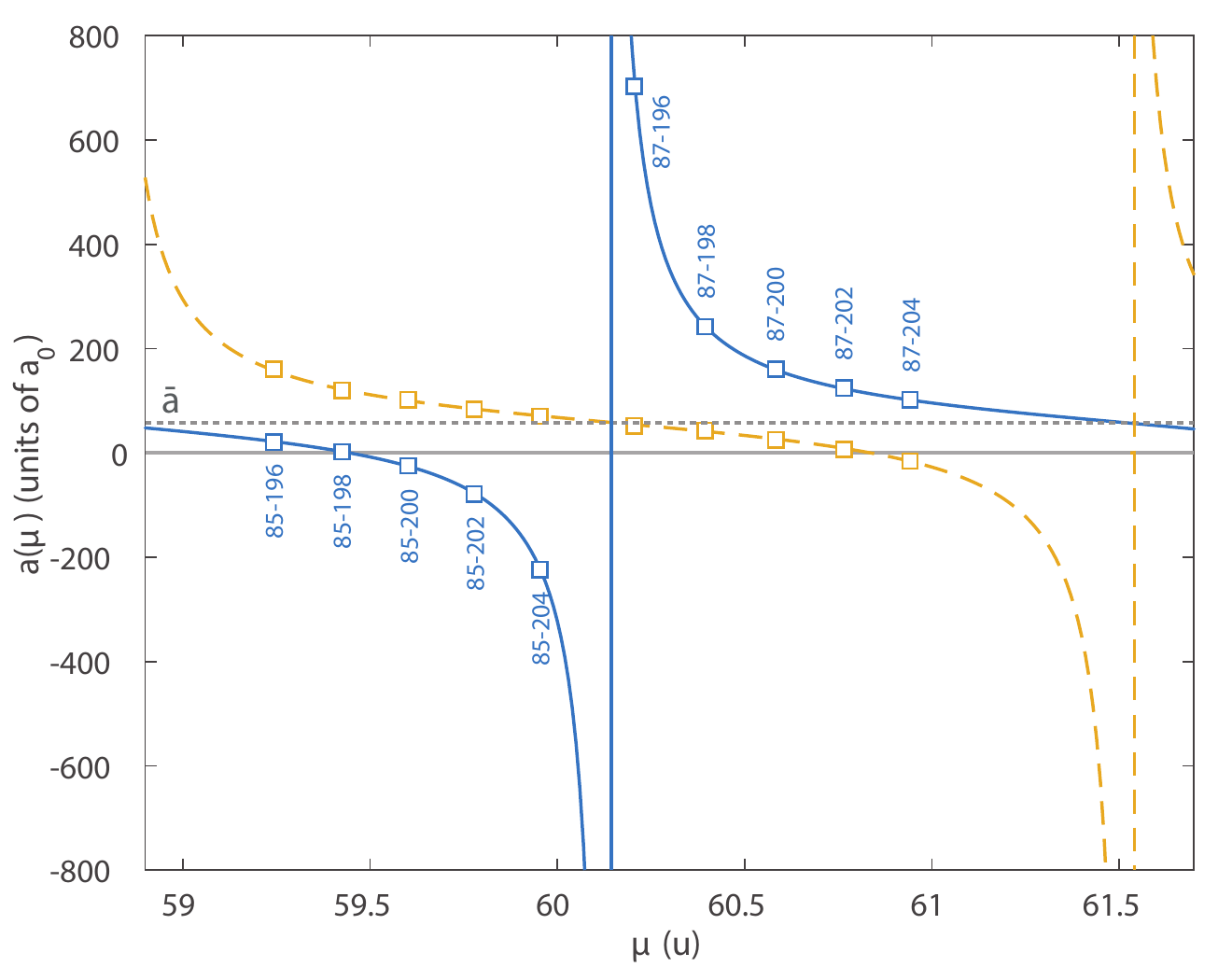}
	\caption{(Color online) Example $s$-wave scattering lengths $a$ for collisions of Rb  and Hg atoms in their respective atomic ground states, $^2$S$_{1/2}$ and $^1$S$_{0}$. Two scenarios are shown: one calculated with the unmodified X~$^2 \Sigma$ \emph{ab initio} potential (blue solid lines) and one where the potential was modified to increase its total WKB phase $\phi$ by $\pi/2$, which corresponds to half a vibrational state (yellow dashed lines). In the former, the available RbHg isotopic combinations would span a wide range of positive and negative scattering lengths with magnitudes both large and small. In the latter case, however, most of the scattering lengths would be positive and of magnitude similar to the mean scattering length $\bar a \approx 57~a_0$ (horizontal dotted grey line). The unusually slow variation of scattering lengths with the reduced mass $\mu$ is due to the shallowness of the RbHg X~$^2 \Sigma$ potential curve (see Fig.~\ref{fig:PEC_HundA}) which only supports 44 bound states (for $^{87}$Rb and $^{202}$Hg). \label{fig:scatlen}}
\end{figure}

While rubidium has two long lived bosonic isotopes, $^{85}$Rb and $^{87}$Rb, the Hg atom features five stable bosons: $^{196}$Hg, $^{198}$Hg, $^{200}$Hg, $^{202}$Hg, and $^{204}$Hg. The $^{202}$Hg isotope is the most abundant (29.74\%). Hence there are ten bosonic isotopomers of RbHg, whose reduced masses 
\begin{equation}
\mu = \left( m_{\rm Rb}^{-1} + m_{\rm Hg}^{-1} \right)^{-1}
\end{equation}
range from 59.24~u to 60.94 u. Selecting an isotopic pair will amount to deciding on the interspecies scattering properties of the mixture. The $p$-wave barrier in RbHg is about 100~$\mu$K high whereas the Doppler temperature for the Hg intercombination line is 30.5~$\mu$K, we expect the interactions in an ultracold Rb+Hg mixture to be dominated by the $s$-wave.

According to the semiclassical approximation \cite{Gribakin1993}, the $s$-wave scattering length 
\begin{equation}
a = \bar a \left[1-\tan \left( \phi-\frac{\pi}{8}\right)\right],
\end{equation}
of a potential with a van der Waals $-C_6/R^6$ long range
is determined by the \emph{mean} scattering length $\bar a = 2^{-\frac{3}{2}} \frac{\Gamma(3/4)}{\Gamma(5/4)}(2\mu C_6/\hbar^2)^{\frac{1}{4}}$ and the zero energy WKB phase integral
\begin{equation}
\phi = \frac{\sqrt{2\mu}}{\hbar}\int_{r_{\rm in}}^{\infty} \sqrt{V(R)}\,dR \,.
\end{equation}
The above integral spans the range between the inner classical turning point $r_{\rm in}$ and infinity. The scattering length $a$ is periodic with respect to the WKB phase $\phi$ (with a period of $\pi$) and shifted by the slowly varying  mean scattering length $\bar a$. For RbHg $\bar a$ varies between 57.20(15)~$a_0$ for the lightest isotopic pair $^{85}$Rb$^{196}$Hg and 57.61(15)~$a_0$ for $^{87}$Rb$^{202}$Hg. The uncertainty in $\bar a$ stems from the accuracy of the $C_6$ coefficient, which we estimate to be about 1~\%. The WKB phase $\phi$ is manifestly proportional to the square root of the reduced mass $\mu$, which we can control by selecting an appropriate isotopic pair. 

Scattering lengths are interrelated with the positions of bound states close to the dissociation limits. The phase integral $\phi$ is determines the number of states $N$ supported by the potential $V(R)$: in fact, within the same approximation
\begin{equation}
N = \left \lfloor \frac{\phi}{\pi} + \frac{3}{8}\right \rfloor\,.
\end{equation}
Singular scattering lengths coincide with bound states located exactly at the dissociation limit.
A large and positive $a$ points to a very weakly bound state just below the dissociation limit.
Finally, scaling the short range potential so that $\phi$ is increased by $\pi$ while $C_6$ is retained amounts to adding one vibrational state with no change to the scattering length.

In RbHg the ground  X~$^2\Sigma$ state potential is very shallow: its depth is only $D_e=404$~cm$^{-1}$ and, for $l=0$, it supports 44 vibrational states. We tentatively estimate the error on $D_e$ to be $\Delta D_e \approx 20$~cm$^{-1}$, or about 5\%. Since the WKB phase $\phi$ depends on the square root of the potential depth, it is determined to within about 2.5\%. This in turn amounts to $\Delta N = 1.09$, a little over one vibrational state. This way we can confirm that N lies between 43 and 45. We can not, alas, predict the actual scattering lengths for all the isotopomers until experimental input, e.g. from a two-color photoassociation experiment \cite{Kitagawa2008, Borkowski2013}, is available.

Figure~\ref{fig:scatlen} shows possible mass scaling behavior of the scattering length $a$ as a function of the reduced mass $\mu$. The ground state potential being shallow, the available bosonic isotopomers span only about two thirds of a scattering length cycle. The singularity in $a$ may fall near one of the isotopomers (blue solid lines in Fig.~\ref{fig:scatlen}). In such case a wide variety of interspecies scattering lengths would be accessible: from large negative to large positive scattering lengths. Only scattering lengths closest to $\bar a$ would be unavailable. For comparison, we also show an opposite, `pessimistic' case (yellow dashed lines in Fig.~\ref{fig:scatlen}), where all scattering lengths have moderate magnitudes and are distributed around $\bar a$. In this case the available scattering lengths would span a range from small negative scattering lengths of tens of $a_0$ to moderate positive of about 150~$a_0$.

\section{Optical Feshbach resonances \label{sec:l_opt}}

\begin{figure}[t]
	\vspace{-1mm}
	\includegraphics[clip, width=0.48\textwidth]{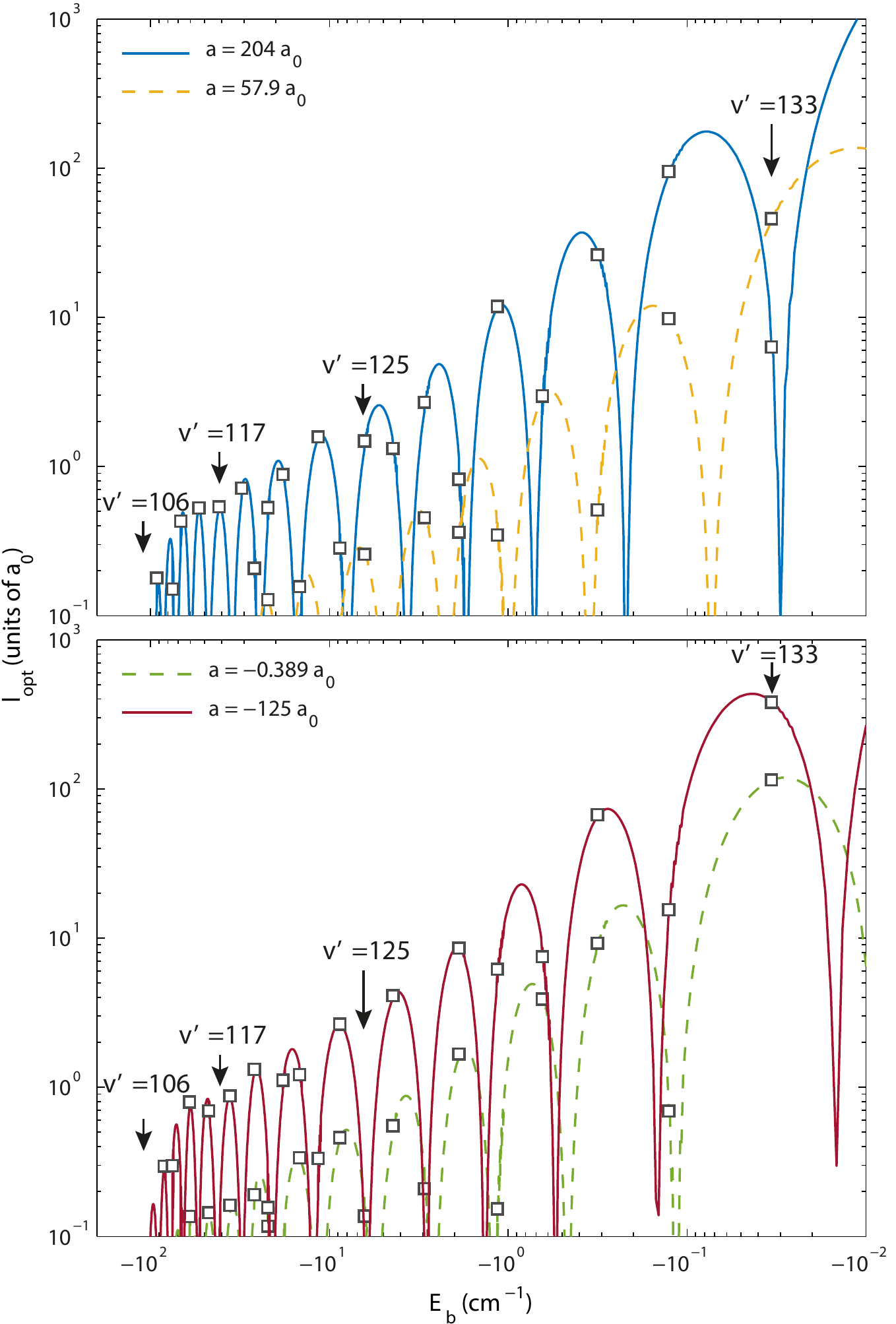}
	\caption{(Color online) Example optical lengths $l_{\rm opt}$ of Feshbach resonances near the Rb atomic transition $^2$S$_{1/2}\,\rightarrow\,^2$P$_{1/2}$ at 795~nm as a function of excited bound state energy $E_b$ shown for characteristic $s$-wave scattering lengths $a$, in order: a large positive scattering length of 204~$a_0$, one close to the mean scattering length ($\bar a \approx 57\,a_0$), one close to zero, and a large negative scattering length. We have marked some of the excited vibrational states $v_{\rm PA}'$ supported by the (2)~$j=1/2\, |\Omega|=1/2$ potential (Fig. \ref{fig:PEC_HundC}) with arrows in order to discuss the products of their optical decay; see Fig.~\ref{fig:fcfs12} and Sec.~\ref{sec:molecule_production} for details.\label{fig:l_opt}}
\end{figure}

Scattering lengths can be manipulated using optical Feshbach resonances (OFRs) \cite{Fedichev1996, Bohn1997, Bohn1999}, where the ground state scattering channel is coupled to an excited molecular bound state by laser radiation. A spectacular example of the use of OFRs was the demonstration of optically controlled collapse of a $^{88}$Sr Bose-Einstein condensate~\cite{Yan2013}. OFRs are by their nature burdened by losses through spontaneous emission from the excited molecular state. While this effect is undesirable in optical \emph{control} of interactions, it is the basis for photoassociation spectroscopy \cite{Jones2006}, which relies on atom loss for the detection of bound state positions. The positions of optical Feshbach resonances from the atomic line are approximately equal to the energies $E_b$ of their respective bound states taken from the appropriate dissociation limit.

Within the isolated resonance theory \cite{Bohn1999, Borkowski2009, Nicholson2015}, in the limit of low collision energies $E \to 0$, the optically modified scattering length near an OFR
\begin{equation}
\alpha(I, \delta) = a + l_{\rm opt}(I)\frac{\delta \gamma}{\delta^2+\frac{\gamma^2}{4}}
\end{equation}
depends dispersively on the laser detuning $\delta$ from the resonance position. The optical length $l_{\rm opt}$ is a resonance strength parameter that determines the maximum change of the scattering length. The magnitudes of the optical lengths are proportional to, and therefore controllable by, the laser intensity $I$; those reported in this paper are calculated for a laser intensity of $I=1\,{\rm W/cm}^2$ and, again, in the limit of low collision energies $E \to 0$. The resonance width $\gamma$ in RbHg is practically equal to the atomic linewidth of the relevant Rb D1 or D2 transition. 

Optical control of scattering lengths comes at a price of two-body photoassociative losses which will impact the trap lifetimes of the dual atomic sample. The time evolution of the atomic densities $n_{\rm Rb}$ and $n_{\rm Hg}$
\begin{equation}
\left\{ \begin{array}{c}
\dot n_{\rm Rb} = -\tilde K_{\rm in} n_{\rm Rb} n_{\rm Hg} - n_{\rm Rb}/\tau_{\rm Rb} \\
\dot n_{\rm Hg} = -\tilde K_{\rm in} n_{\rm Rb} n_{\rm Hg} - n_{\rm Hg}/\tau_{\rm Hg}
\end{array}
\right.	
\end{equation}
may be severely modified if the two-body inelastic rate $\tilde K_{in}$ multiplied by appropriate atomic densities takes over the one-body lifetimes $\tau_{\rm Rb}$ and $\tau_{\rm Hg}$. For thermal atoms $\tilde K_{\rm in}$ may be calculated as an appropriate Boltzmann average of the kinetic energy dependent loss rate \cite{Nicholson2015}:
\begin{equation}
K_{\rm in}(E) = \frac{4\pi\hbar}{\mu}\frac{\gamma^2 l_{\rm opt}}{(\delta+E/\hbar)^2 + \frac{\gamma^2}{4}(1+2k l_{\rm opt})^2} \label{eq:K_in}
\end{equation}
which in itself is a Lorentz curve shifted to the red by the kinetic energy $E$ of the colliding atoms. The thermal averaging leads to an asymmetric lineshape that may be calculated numerically \cite{Ciurylo2004} or approximated with the formulas of Ref.~\cite{Jones1999}.

The optical length $l_{\rm opt}$ incorporates the molecular physics involved in the optical Feshbach resonance process. An $s$-wave collision of Rb and Hg atoms at a kinetic energy $E$ may be described by an energy-normalized scattering wavefunction $\Psi(R, E) \xrightarrow{R\to \infty} \sqrt{2\mu/\pi\hbar^2 k} \sin(kR+\eta)$, where the wavenumber $k=\sqrt{2\mu E}/\hbar$ and $\eta$ is a phase shift introduced by the short range ground state interaction potential. The radial motion in the excited bound state is described by the unit-normalized wavefunction $\Psi'(R)$. Both can be calculated by solving appropriate single channel Hund's case (c) Schr\"odinger equations. Assuming that the transition dipole moment is constant for large internuclear distances (it is -- see Fig.~\ref{fig:matrixelements}c), the optical length can be given by a Franck-Condon factor
\begin{equation}
l_{\rm opt} = \frac{3}{16\pi k} \frac{I \lambda^3}{c} f_{\rm rot} \left|\int_{0}^{\infty} \Psi(R, E)\Psi'(R) dR \right|^2,
\end{equation} 
where $\lambda$ is the transition wavelength, $I$ is the laser intensity. The H\"onl-London factor $f_{\rm rot}$ accounts for the rotational couplings. For $s$-wave transitions to the excited $j$=3/2 states $f_{\rm rot}$=1/2 (because these states mix different rotational states), whereas for $j$=1/2 $f_{\rm rot}$=1 (because these states do not); see Sec.~\ref{sec:couplings} for details. The polarization-independent $f_{\rm rot}$ factors are valid when no external fields fix the atomic $z$ axis and low light intensities are used \cite{Borkowski2009}. If, however, such fields are present, the projection $M$ of the total angular momentum on the $z$ axis also has to be considered. By virtue of the Wigner-Eckart theorem, the polarization-dependent rotational factor 
\begin{equation}
	f_{\rm rot} (M, q) = f_{\rm rot} \left| \left<J_g K M_g q| J_e M_e \right> \right|^2 \,.
\end{equation}
For dipole transitions $K = 1$ while $q = -1,0,+1$ denotes laser polarizations $\sigma^-$, $\pi$ and $\sigma^+$, respectively. 

\begin{figure}[t]
	\includegraphics[width=0.5\textwidth]{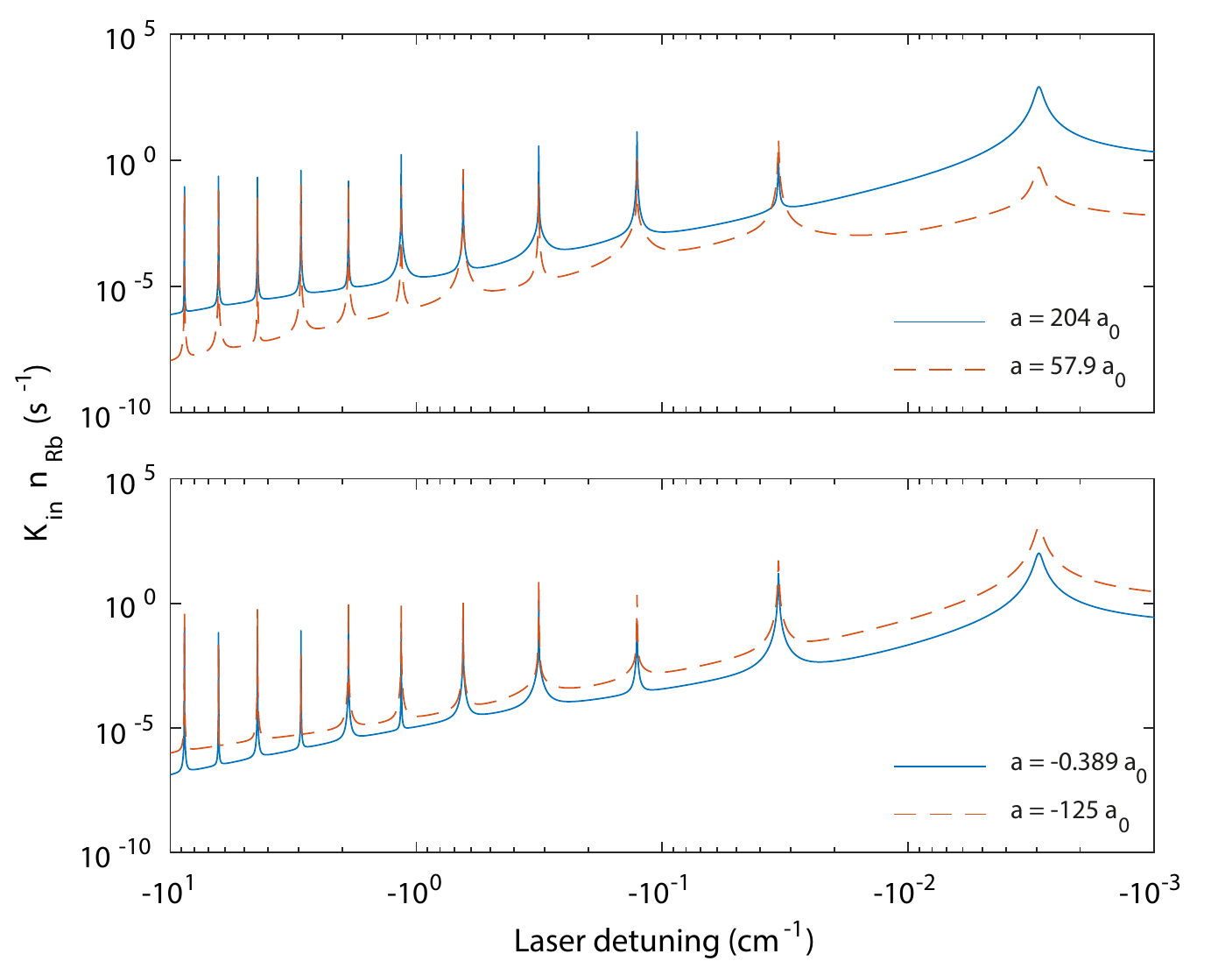}
	\caption{(Color online) Thermally averaged photoassociation-induced Hg atom loss rates $\tilde K_{\rm in} n_{\rm Rb}$ near the 795~nm transition to the RbHg (2)~j=1/2~$|\Omega|$=1/2 state calculated using Eq.~(\ref{eq:K_in}) for sample temperature of $T=100\, \mu K$, laser intensity $I=500~{\rm W}/{\rm cm^2}$ and the number density of the Rb cloud of $n_{\rm Rb} = 10^9\,{\rm cm}^{-3}$. For many photoassociation lines the two-body loss rates reach $1\,{\rm s}^{-1}$ which is more than the one-body losses of the Hg MOT reported in Ref.~\cite{Witkowski2017}. This will enable direct detection of heteronuclear photoassociation by monitoring the steady-state populations in a two-species MOT much like in the previous RbYb experiment \cite{Nemitz2009}. \label{fig:spectra}}
\end{figure}

\begin{table}
	\caption{Example routes to the RbHg ground (0,0,0) state. The entry point for each route is photoassociation to (2)~j=1/2,~$|\Omega|$=1/2 vibrational state $v'$, the strength of which (at a laser intensity of 1~W/cm$^2$) is described by its optical length $l_{\rm opt}$. The optical lengths are strongly dependent on the scattering length $a$. The excited molecules decay to the ground vibrational state $v$ at an efficiency determined by the Franck-Condon factor $f(v,v'_{\rm PA})$. The ground rovibrational state may be transferred by a STIRAP process through the intermediate state $v'$ using lasers of wavelengths $\lambda_1$ and $\lambda_2$ with transition probabilities $f(v,v')$ and $f(0,v')$, respectively. \label{tab:routes}}
	\begin{ruledtabular}
		\begin{tabular}{l r r r r}
			& Route 1 & Route 2 & Route 3 & Route 4 \\
			\hline
			$v'_{\rm PA}$                      & 133     & 125     & 117     & 106      \\ 
			$\lambda_{\rm PA}$ (nm)           & 795.0   & 795.4   & 797.3   & 802.7   \\
			$l_{\rm opt}$ ($a = 204\, a_0$)   & 6.33   & 1.49    & 0.016   & 0.047   \\
			$l_{\rm opt}$ ($a = 57.9\, a_0$)  & 45.8     & 0.25   & 0.040   & 0.0066  \\
			$l_{\rm opt}$ ($a = -0.4\, a_0$)  & 115     & 0.13    & 0.16    & 0.0084    \\
			$l_{\rm opt}$ ($a = -125\, a_0$)  & 381 &   0.025 &   0.88 &   0.027  \\
			$v$                               & 42      & 37      & 33      & 25      \\
			$f(v,v'_{\rm PA})$                 & 0.59    & 0.27    & 0.23    & 0.36    \\	
			$v'$                              & 41      & 41      & 41      & 39      \\
			$f(v,v')$                         & 2.1$\times$10$^{-5}$ & 5.1$\times$10$^{-4}$ & 0.0016 & 0.0043\\
			$f(0,v')$                        & 0.067   & 0.067   & 0.067   & 0.068  \\
			$\lambda_1$ (nm) & 1073.9 & 1073.7 & 1072.8 & 1085.9 \\
			$\lambda_2$ (nm) & 1030.4 & 1030.4 & 1030.4 & 1046.9 \\
		\end{tabular}
	\end{ruledtabular}
\end{table}

The photoassociative loss of atoms is, like the optical shift to the scattering length, proportional to the optical length $l_{\rm opt}$. Their dependence on the laser detuning, however, is different: for large detunings $\delta$, the modification to the scattering length diminishes as $1/\delta$, whereas the photoassociative losses scale as $1/\delta^2$. For this reason it is commonly recommended \cite{Ciurylo2005} that large detunings and laser intensities be used to minimize inelastic losses while maintaining control over the scattering length.

The behavior of optical lengths of resonances close to the atomic limit for the RbHg excited states considered in this work is qualitatively very similar; therefore we will only take the 795~nm transitions to the RbHg (2)~j=1/2~$|\Omega|$=1/2 state as an example. Again, we stress that the current \emph{ab initio} calculations do not make it possible to accurately predict the positions of bound states close to the dissociation limit. The vibrational spacings, however, being dependent on the $C_6$ coefficient, are correct. For this reason in Fig.~\ref{fig:l_opt} we show optical lengths as a function of the (unknown) bound state position $E_b$ with an example series of bound states marked as squares which was calculated for an unmodified potential curve and a reduced mass matching the $^{87}$Rb$^{202}$Hg isotopologue. We have also marked four example vibrational states as `entry points' for possible routes to ground state RbHg molecules discussed in Sec.~\ref{sec:molecule_production}.

The positions of optical Feshbach resonances in a given potential can be fixed by finding at least one resonance in experiment. The lower bounds for the top two energy bins for the (2)~j=1/2~$|\Omega|$=1/2 state are $E_1 = 232$~MHz and $E_2=1596$~MHz meaning that the top two resonances are within reach of a single acousto-optic modulator from the atomic line. Given the large optical lengths for bound states very close to the dissociation limit, we expect that optical Feshbach resonances in this system could be detected by simply monitoring the fluorescence of the Hg MOT, much like in the case of the Rb+Yb experiment reported in Ref.~\cite{Nemitz2009}. This is further corroborated by our simulations of the photoassociation spectra in Fig.~\ref{fig:spectra}. For a realistic number density of the Rb cloud of $n_{\rm Rb}=10^9\,{\rm cm}^{-3}$ and a modest laser intensity of $I=500\,{\rm W/cm}^2$ the photoassociative loss rate for the Hg atoms, $\tilde K_{\rm in} n_{\rm Rb}$, easily exceeds the one-body loss rate of $1/\tau_{\rm Hg}\approx 0.4\,s^{-1}$.

The choice of a resonance for optical control of the scattering length will be driven by the magnitude of its optical length and possible photoassociative losses. Optical lengths depend primarily on the scattering length $a$ and the position of the relevant excited bound state. Bound states closer to the dissociation limit yield higher magnitudes of the optical length because their bound state wavefunctions have better overlap with the scattering wavefunction in the ground state. Large magnitudes of the scattering length $a$ enhance the optical lengths: it is a general property of cold $s$-wave scattering that the amplitude of the short range part of the scattering wavefunction grows significantly when $|a| \gg \bar a$ \cite{Chin2010}. Finally, $l_{\rm opt}$ has nodes stemming from the oscillatory character of the ground and excited wavefunctions. For example, the optical length for the $v'=133$ line when the scattering length $a=204~a_0$ is seriously diminished by such a node even though this does not happen for the other scattering lengths. Such a problem can be worked around by selecting either a different isotopic pair (which influences $a$) or using a neighbor state, like $v'=132$.

\begin{figure*}
	\centering
	\includegraphics[width=1.0\textwidth]{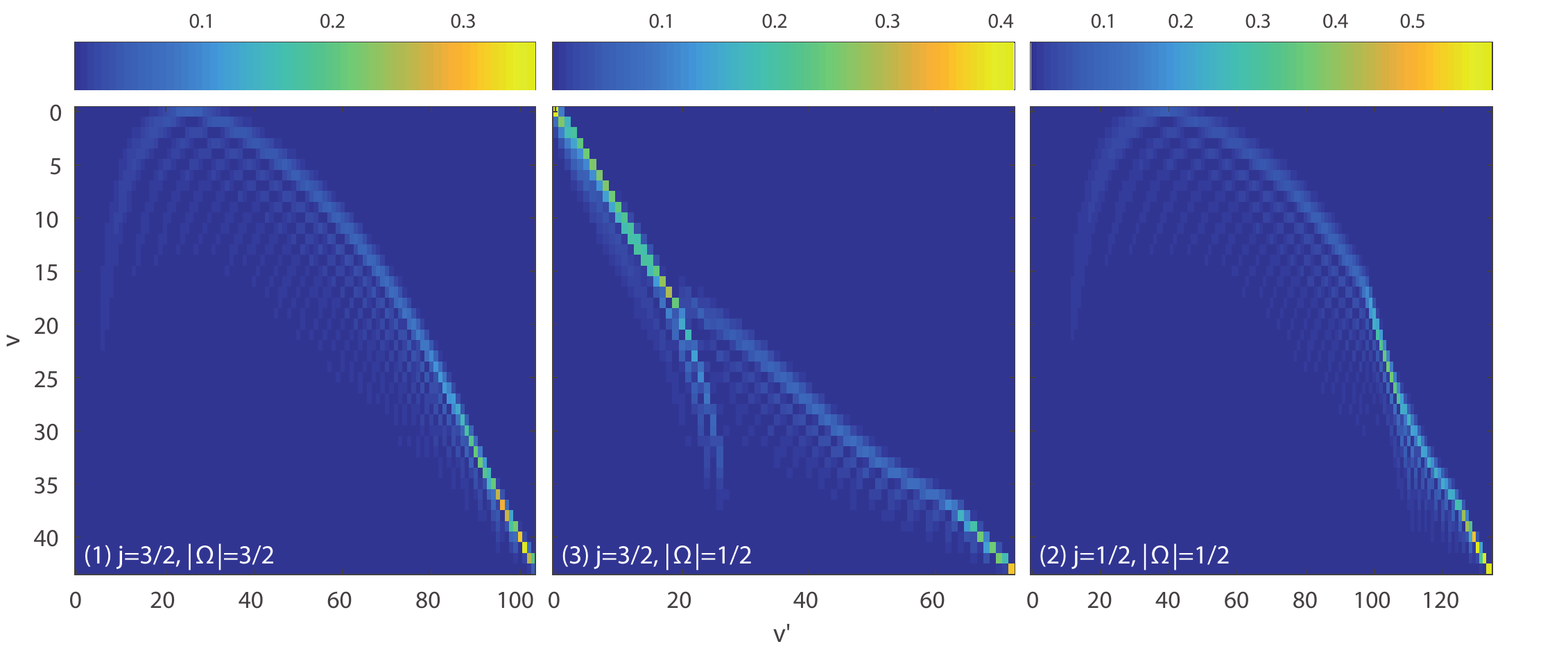}
	\caption{(Color online) Effective Franck-Condon factors calculated for transitions between vibrational states $v'$ of appropriate excited electronic potentials near the $^2$P+$^1$S asymptote and vibrational states $v$ in the X~$^2$S+$^1$S ground state of the RbHg molecule. The internuclear distance variability of the transition dipole moment was taken into account, see Eq.~(\ref{eq:fcf}). It is interesting to note that the Franck-Condon factors for the $(3)\, j=3/2,\, |\Omega|=1/2$~$\rightarrow$~$(1)\,j=1/2,\,|\Omega|=1/2$ transition are unusually diagonal for the lowest 20 vibrational states -- which may be the result of both curves having similar equilibrium distances $R_e$. \label{fig:fcfmaps}}
\end{figure*}

Due to the $C_6/R^6$ asymptotics in the RbHg interactions, the OFRs are sparsely distributed, theoretically making it possible to detune the OFR laser further from resonance. This is in opposition to the case of Rb+Rb OFRs, where the $C_3/R^3$ dipole interaction supports many resonances close to the dissociation limit leading to considerable photoassociative losses \cite{Theis2004}. Operating an OFR laser very close to the Rb 795~nm line will likely cause significant heating of the atomic sample and strong Rb-Rb photoassociative losses: close to the dissociation limit the photoassociation spectrum of Rb is practically a continuum \cite{Miller1993}. To avoid this, it might be necessary to choose a Rb+Hg OFR at larger detunings from the atomic resonance, even at the expense of a smaller optical length. There are also other possible workarounds. Aside from obviously lowering the density of the Rb cloud which would quadratically reduce the Rb+Rb photoassociation rate, a much more interesting idea could be to manipulate the photoassociation rate for Rb by magnetically changing its scattering length \cite{Pellegrini2008}. Since optical lengths have nodes whose positions depend on the scattering length, it may be possible to engineer a node in Rb+Rb photoassociation at detunings where a Rb+Hg OFR laser would operate.

\section{Prospects for ground state molecule production \label{sec:molecule_production} }

Photoassociative loss of atoms is usually due to the spontaneous emission from the excited molecular state. The excited molecule may decay to any ground state energy level below its original energy. If the target state lies in the continuum above the ground state asymptote, the atoms are no longer bound and gain kinetic energy; in RbHg most will be absorbed by the lighter Rb atom. If the kinetic energy is much larger than the trap depth, one or both atoms may be ejected from the trap and lost forever. If, however, the kinetic energy of the target continuum state is low, the atoms remain in the trap and may take part in another photoassociation cycle.
Finally, the excited molecule may decay to a bound state in the ground electronic state. The likelihood of forming a ground state molecule in vibrational and rotational states $v$ and $J_g$ is equal to the bound-bound Franck-Condon factor 
\begin{equation}
f(v,v') =  f_{\rm rot} \left|\int_0^\infty \Psi(v, R) \frac{d(R)}{d_{\rm at}}  \Psi'(v', R) dR \right|^2 \label{eq:fcf}
\end{equation}
between, respectively, the ground state and excited state wavefunctions $\Psi(v, R)$ and $\Psi'(v', R)$ representing the vibrational states $v$ and $v'$. The H\"onl-London rotational factor $f_{\rm rot}$ represents the effect of the rotational couplings described in Sec.~\ref{sec:couplings}. We have also included the dependence of the transition dipole moments $d(R)$ on the internuclear distance $R$ with respect to the asymptotic (atomic) dipole moment of Rb, $d_{\rm at}$. The appropriate transition dipole moments are shown in Fig.~\ref{fig:matrixelements}.

The Franck-Condon factors for the 780~nm transitions from the (1)~j=3/2,~$|\Omega|$=3/2 (Fig.~\ref{fig:fcfmaps}, left panel) and the 795~nm transitions from (2)~j=1/2,~$|\Omega|$=1/2 (Fig.~\ref{fig:fcfmaps}, right panel) states are qualitatively very similar: the shapes of both curves are determined primarily by the Hund's case (a) 1~$^2\Pi$ potential. The Franck-Condon factors are significantly non-diagonal. From the point of view of forming ground state molecules in the rovibrational ground state, the broad maximum of $f(v,v')$ for $v'\approx 20 \ldots 40$ will be very useful for the final STIRAP process. On the other hand, the Franck-Condon factors for the 780~nm transitions from the (3)~j=3/2~$|\Omega|$=1/2 state (Fig.~\ref{fig:fcfmaps}, center panel) have an unusually diagonal character for the lowest 20 or so vibrational states which can be explained by the similarity of equilibrium distances, $R_e$=7.61~$a_0$~and 7.82~$a_0$, of the relevant excited and ground state potential curves.

The lowest rotational state for $j=3/2$ electronic states is $J=3/2$ and, as shown in Sec.~\ref{sec:couplings}, is a mixture of $l=0$ and $l=2$ state. The rotational factors for transitions to the $s$-wave ground state are therefore $f_{\rm rot}=1/2$. In other words, half of the photoassociated molecules decay to the $s$-wave, and the other half to the $d$-wave. On the other hand, the excited $j=1/2$, $J=1/2$ states produced via photoassociation from the $s$-wave are purely $l=0$. Thus, for the remainder of this work, we will consider the production of ground state RbHg molecules using the Rb 795~nm $^2S_{1/2}\,\rightarrow\,^2P_{1/2}$ transition.

\begin{figure}[t]
	\vspace{0.5mm}
	\includegraphics[width=0.498\textwidth]{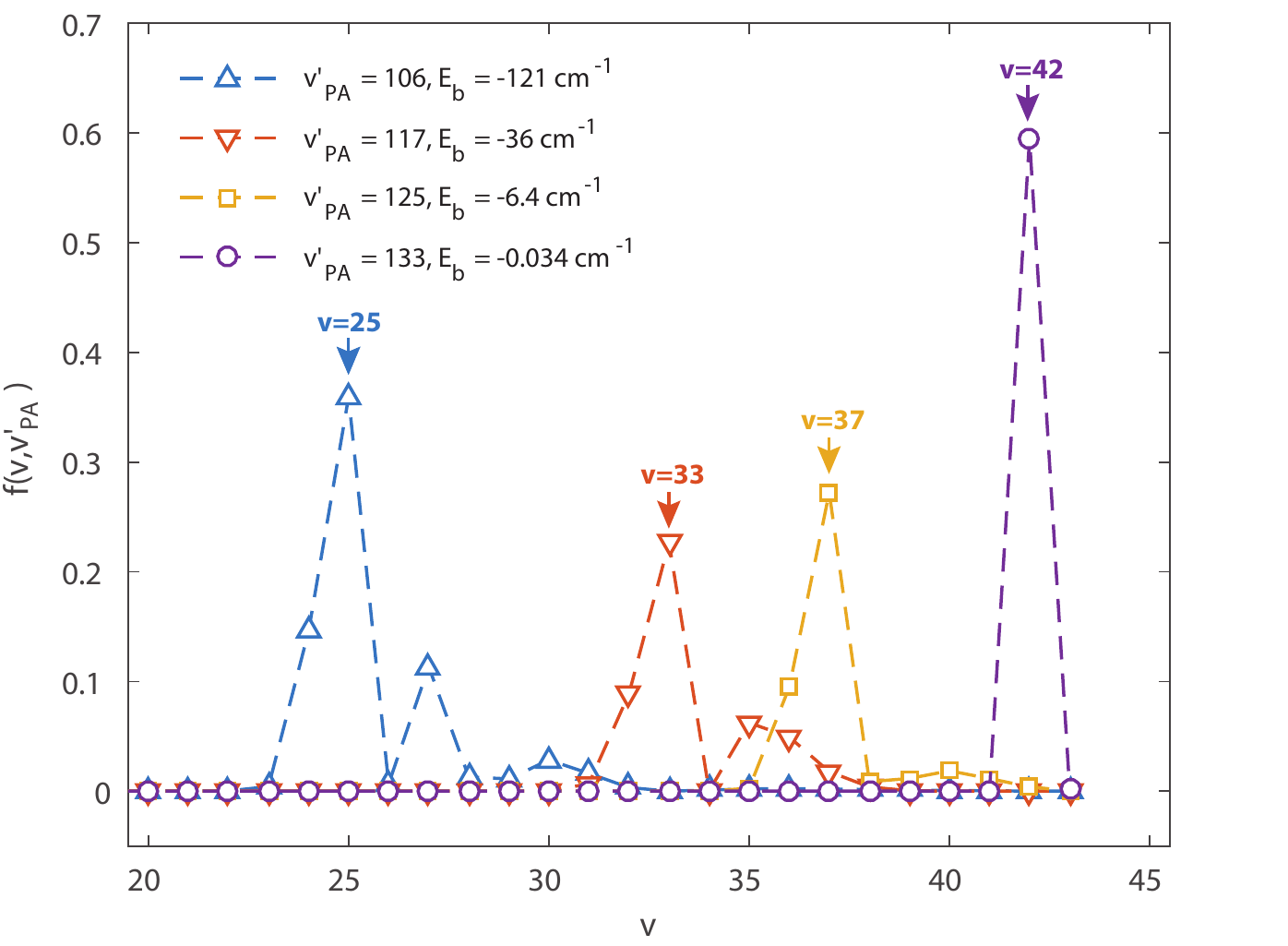}
	\caption{(Color online) Effective Franck-Condon factors for transitions between selected (2)~j=1/2~$|\Omega|$=1/2 vibrational states $v'$ and ground vibrational states $v$. For many of the excited state vibrational levels $v'$ the product states are dominated by one ground energy level $v$. For example, photoassociation of Rb and Hg atoms to the $v'_{\rm PA} = 125$ state could, through spontaneous emission,  produce ground state RbHg molecules in the $v=37,\, l=0$ rovibrational state at an efficiency of over 25\%. \label{fig:fcfs12}}
\end{figure}

Figure~\ref{fig:fcfs12} shows the probabilities of forming ground state molecules in the rovibrational state $v,\,l=0$ by spontaneous emission from excited states $v'$=133,~125,~117~and~106. These will constitute the entry points for our example routes to the rovibrational RbHg ground state collected in Table~\ref{tab:routes}. The respective photoassociative optical lengths are also marked in Figure~\ref{fig:l_opt}. The Franck-Condon factors have favorable properties. One can select an excited state $v'_{\rm PA}$ that produces practically any desired ground vibrational state $v \gtrapprox 25$ at an efficiency of at least 20\%. For example, if the desired ground vibrational state is $v=37$ (Route 2 in Table~\ref{tab:routes}), then one could perform photoassociation using the excited $v'_{\rm PA}=125$ state at $-6.4\,{\rm cm}^{-1}$. The molecular production efficiency would be about 27\%.

Similarly to the case of optical control of interactions, the selection of an appropriate photoassociation line will be influenced by the optical length $l_{\rm opt}$. The choice of an appropriate isotopic pair is important:  scattering lengths of large magnitude (positive or negative) give larger optical lengths than those close to the the mean scattering length $\bar a$. Bound states close to the dissociation limit are much easier to photoassociate, because of the more favorable free-bound Franck-Condon factors. These, however, have the disadvantage that the ground state molecules will be produced in vibrational states closer to the dissociation limit which will have a negative impact on the Franck-Condon factors of the transitions used in the later STIRAP process. On the other hand, the loss of rubidium atoms is no longer that much of a concern. In typical experiments involving Rb and divalent atoms, the Rb population is about an order of magnitude larger than the other (see e.g. \cite{Witkowski2017, Nemitz2009}). Performing photoassociation using deeper lying vibrational states may, however, be beneficial despite their smaller optical lengths (which can partially be compensated for using higher laser intensities), because of the easier STIRAP process later.

\section{STIRAP}

\begin{figure}[t]
	\includegraphics[width=0.505\textwidth]{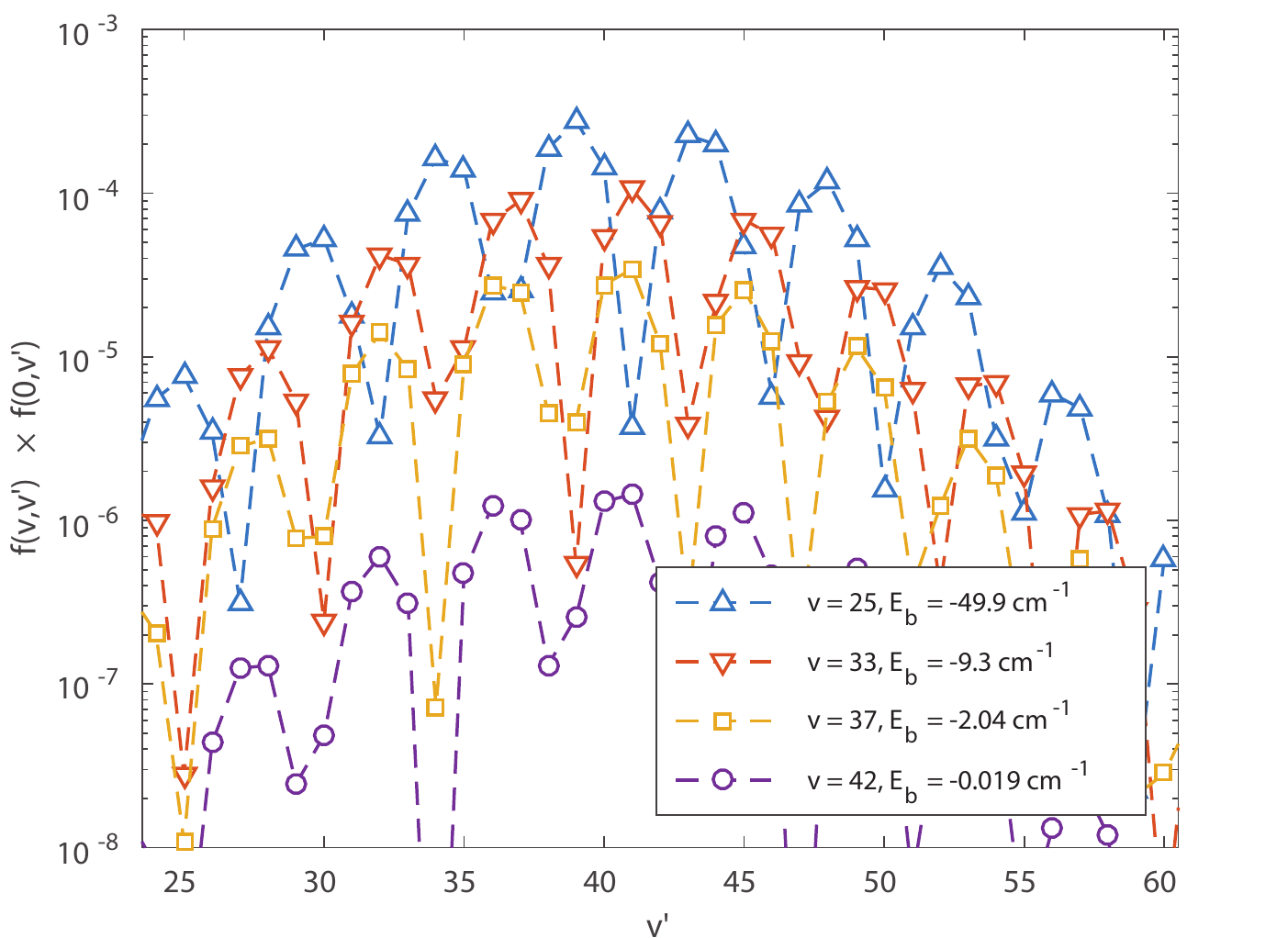}
	\caption{(Color online) Products of Franck-Condon factors describing the transition probabilities in STIRAP processes between a starting ground vibrational level $v$ of energy $E_b$, an intermediate vibrational level $v'$ in the (2)~j=1/2~$|\Omega|$=1/2 excited electronic state and the rovibrational ground state. \label{fig:doublefcfs}}
\end{figure}

The weakly bound ground state molecules can be transferred to the rovibrational ground state using the STIRAP technique \cite{Bergmann1998}. Following a similar analysis of the RbSr system of Chen~\emph{et al.} \cite{Chen2014}, in Fig.~\ref{fig:doublefcfs} we show the products of Franck-Condon factors for the two transitions between the initial ground vibrational states $v$ and target rovibrational ground state as a function of the intermediate excited vibrational state~$v'$. A large Franck-Condon factor product $f(v,v')\times f(0,v')$ means that lower laser powers can be used during the STIRAP process. 

For the example routes the calculated products are mostly favorable and range between $1.4\times10^{-6}$ (Route 1, $v=42$ and $v'=41$) and $2.9\times10^{-4}$ (Route 4, $v=25$ and $v'=39$). In comparison, numerical simulations for the RbSr system found that values of on the order of about $3\times 10^{-5}$ are sufficient to achieve STIRAP round trip efficiencies of about 60\%. Routes 2,~3,~and~4 have at least as favorable Franck-Condon factors. Generally, the more deeply bound initial vibrational state $v$, the easier the STIRAP process will be. Again, the selection of a route to ground state molecules will be a compromise between photoassociation efficiency (which favors excited states close to the dissociation limit), round-trip STIRAP efficiency (which is better if the initial state ground vibrational state is deeply bound) and the available lasers. Luckily, diode lasers are available for both photoassociation (795~nm -- 803~nm) and the STIRAP pump (1073.9~nm -- 1085.9~nm) and dump (1030.4~nm -- 1046.9~nm) wavelengths. We expect that a practical realization would use a route similar to either Route 2 or 3, which avoids the disadvantages related to either too small optical lengths (Route 4) or unfavorable Franck-Condon factors for the STIRAP transfer (Route 1).

\section{Summary and conclusion \label{sec:summary}}

Stimulated by the experimental progress in trapping the ultracold mixture of Rb and Hg atoms~\cite{Witkowski2017}, we have explored theoretically the prospects for photoassociation of these two atoms near the Rb $^2$S$\rightarrow$$^2$P excitation thresholds, and the possibility of optical manipulation of scattering properties of the ultracold Rb+Hg gas. We have also studied the prospects  for the formation of RbHg molecules in the rovibrational ground state.

We have carried out state-of-the-art {\em ab initio } calculations of the potential energy curves for the RbHg system using the coupled-clusters method for ground and excited states. The RbHg interactions are unusually weak given that both atoms are metals: the X~$^2$$\Sigma$ ground state potential 
is only 404 cm$^{-1}$ deep with a potential minimum at about 7.8 $a_0$ and it supports 44 vibrational states. This is very shallow compared to RbSr or RbYb molecules, which bear a similar electronic configuration. The excited states are also quite peculiar as the potential shapes differ very strongly: the 2~$^2\Sigma^+$ excited state is very shallow and has an equilibrium distance similar to that of the ground state. On the other hand the potential well of the 1~$^2\Pi$ state is quite deep, similar to other alkali-metal+divalent~atom systems.

The RbHg system offers prospects for optical interspecies interaction control via optical Feshbach resonances near the Rb $^2$S$\rightarrow$$^2$P transition. The optical lengths at 1~W/cm$^2$ laser intensities, depending on the background interspecies scattering length, can be as large as tens of $a_0$ even at red-detunings close to 1~cm$^{-1}$ raising hope for low-loss control of interactions at moderate laser intensities. Since the Hg atom offers several stable isotopes it may be possible to use mass-tuning to select a favorable isotopic mixture to employ OFR. The main difficulty in the implementation of OFRs in this system is the photoassociative loss of Rb atoms due to the dense spectrum of Rb-Rb resonances. These could be circumvented by preparing the mixture in a 3D optical lattice or by lowering Rb atoms density in the trap, or by reducing the rubidium photoassociation rate using a Rb magnetic Feshbach resonance. We note that OFRs may be the only technique available for control of scattering lengths in this system as we expect that magnetic Feshbach resonances created by the modification of the Rb hyperfine interaction \cite{Zuchowski2010, Brue2012} would be very weak because the charge transfer between Rb and Hg atoms is small as evidenced by the relatively marginal dipole moment of RbHg.

We have identified possible paths for rovibrational ground state RbHg molecule formation. In the first step, weakly bound ground state molecules are created by spontaneous emission as a byproduct of photoassociation. We find that not only molecules in the top vibrational states close to the dissociation limit could be produced at an efficiency of about 50\%, but it is also possible to select a 795~nm photoassociation resonance that deeply bound $v \approx 25\ldots35$ states can be made at efficiencies $\sim 20$\%. The second stage involves a STIRAP transfer from the weakly bound ground vibrational level to the rovibrational ground state. We find that  it is in many cases possible to select an intermediate excited state where the product of the two relevant Franck-Condon factors is sufficient to achieve high STIRAP efficiencies~\cite{Chen2014}. The favorable Franck-Condon factors are thanks to the much shorter equilibrium distance of the 1~$^2\Pi$ excited state compared to the ground state. We note that this scheme does not rely on having an optical lattice and can be conducted using commercially available diode lasers. We also note that the favorable Franck-Condon factors can be found without relying on enhancement mechanisms \cite{Dion2001, Pechkis2007} due to resonant coupling between $j=1/2$ and $j=3/2$ states. In principle, however, once experimental input is available, the positions of enhanced molecular states could be predicted and utilized in the production of tightly bound RbHg molecules.

In the near future we plan to perform photoassociation spectroscopy in a dual MOT of Rb and Hg by monitoring the fluorescence of the Hg cloud using a  photomultiplier. Optical trapping of the Hg atoms in the optical trap is challenging due to its low dynamic polarizability (about 50 a.u. for a 1.5 $\mu$m laser), but possible, given the Doppler limit for Hg intercombination line is low (31 $\mu$K). Long term goals include high precision spectroscopy and molecule formation for the purposes of the search for new physics~\cite{Meyer2009}.

\acknowledgements
We would like to acknowledge support from National Science Center of Poland (NCN) grants No. 2012/07/B/ST4/01347 (MK), 2012/07/B/ST2/00235 (PSZ, RMR and RC) and  2014/13/N/ST2/02591 (MB). We also are grateful to Fundation for Polish Science for funding of Homing Plus project no. 2011-3/14 which was co-financed by the EU European Regional Development Fund.
Support has been received from the project EMPIR 15SIB03 OC18. This project has received funding from the EMPIR programme co-financed by the Participating States and from the European Union’s Horizon 2020 research and innovation programme.
Calculations have been carried out using resources provided by the Wroclaw Centre for Networking and Supercomputing (http://wcss.pl), grant Nos. 218 (PSZ) and 353 (MB). This work is part of the ongoing research program of the National Laboratory FAMO in Toru\'n, Poland.

\bibliography{HgRb,library} %

\begin{thebibliography}{85}%
\makeatletter
\providecommand \@ifxundefined [1]{%
 \@ifx{#1\undefined}
}%
\providecommand \@ifnum [1]{%
 \ifnum #1\expandafter \@firstoftwo
 \else \expandafter \@secondoftwo
 \fi
}%
\providecommand \@ifx [1]{%
 \ifx #1\expandafter \@firstoftwo
 \else \expandafter \@secondoftwo
 \fi
}%
\providecommand \natexlab [1]{#1}%
\providecommand \enquote  [1]{``#1''}%
\providecommand \bibnamefont  [1]{#1}%
\providecommand \bibfnamefont [1]{#1}%
\providecommand \citenamefont [1]{#1}%
\providecommand \href@noop [0]{\@secondoftwo}%
\providecommand \href [0]{\begingroup \@sanitize@url \@href}%
\providecommand \@href[1]{\@@startlink{#1}\@@href}%
\providecommand \@@href[1]{\endgroup#1\@@endlink}%
\providecommand \@sanitize@url [0]{\catcode `\\12\catcode `\$12\catcode
  `\&12\catcode `\#12\catcode `\^12\catcode `\_12\catcode `\%12\relax}%
\providecommand \@@startlink[1]{}%
\providecommand \@@endlink[0]{}%
\providecommand \url  [0]{\begingroup\@sanitize@url \@url }%
\providecommand \@url [1]{\endgroup\@href {#1}{\urlprefix }}%
\providecommand \urlprefix  [0]{URL }%
\providecommand \Eprint [0]{\href }%
\providecommand \doibase [0]{http://dx.doi.org/}%
\providecommand \selectlanguage [0]{\@gobble}%
\providecommand \bibinfo  [0]{\@secondoftwo}%
\providecommand \bibfield  [0]{\@secondoftwo}%
\providecommand \translation [1]{[#1]}%
\providecommand \BibitemOpen [0]{}%
\providecommand \bibitemStop [0]{}%
\providecommand \bibitemNoStop [0]{.\EOS\space}%
\providecommand \EOS [0]{\spacefactor3000\relax}%
\providecommand \BibitemShut  [1]{\csname bibitem#1\endcsname}%
\let\auto@bib@innerbib\@empty
\bibitem [{\citenamefont {Moses}\ \emph {et~al.}(2015)\citenamefont {Moses},
  \citenamefont {Covey}, \citenamefont {Miecnikowski}, \citenamefont {Yan},
  \citenamefont {Gadway}, \citenamefont {Ye},\ and\ \citenamefont
  {Jin}}]{Moses2015}%
  \BibitemOpen
  \bibfield  {author} {\bibinfo {author} {\bibfnamefont {S.~A.}\ \bibnamefont
  {Moses}}, \bibinfo {author} {\bibfnamefont {J.~P.}\ \bibnamefont {Covey}},
  \bibinfo {author} {\bibfnamefont {M.~T.}\ \bibnamefont {Miecnikowski}},
  \bibinfo {author} {\bibfnamefont {B.}~\bibnamefont {Yan}}, \bibinfo {author}
  {\bibfnamefont {B.}~\bibnamefont {Gadway}}, \bibinfo {author} {\bibfnamefont
  {J.}~\bibnamefont {Ye}}, \ and\ \bibinfo {author} {\bibfnamefont {D.~S.}\
  \bibnamefont {Jin}},\ }\href
  {http://science.sciencemag.org/content/350/6261/659} {\bibfield  {journal}
  {\bibinfo  {journal} {Science}\ }\textbf {\bibinfo {volume} {350}} (\bibinfo
  {year} {2015})}\BibitemShut {NoStop}%
\bibitem [{\citenamefont {Baranov}\ \emph {et~al.}(2012)\citenamefont
  {Baranov}, \citenamefont {Dalmonte}, \citenamefont {Pupillo},\ and\
  \citenamefont {Zoller}}]{Baranov2012}%
  \BibitemOpen
  \bibfield  {author} {\bibinfo {author} {\bibfnamefont {M.~A.}\ \bibnamefont
  {Baranov}}, \bibinfo {author} {\bibfnamefont {M.}~\bibnamefont {Dalmonte}},
  \bibinfo {author} {\bibfnamefont {G.}~\bibnamefont {Pupillo}}, \ and\
  \bibinfo {author} {\bibfnamefont {P.}~\bibnamefont {Zoller}},\ }\href
  {\doibase 10.1021/cr2003568} {\bibfield  {journal} {\bibinfo  {journal}
  {Chem. Rev.}\ }\textbf {\bibinfo {volume} {112}},\ \bibinfo {pages} {5012}
  (\bibinfo {year} {2012})}\BibitemShut {NoStop}%
\bibitem [{\citenamefont {B{\"{u}}chler}\ \emph {et~al.}(2007)\citenamefont
  {B{\"{u}}chler}, \citenamefont {Demler}, \citenamefont {Lukin}, \citenamefont
  {Micheli}, \citenamefont {Prokof'ev}, \citenamefont {Pupillo},\ and\
  \citenamefont {Zoller}}]{Buechler2007}%
  \BibitemOpen
  \bibfield  {author} {\bibinfo {author} {\bibfnamefont {H.~P.}\ \bibnamefont
  {B{\"{u}}chler}}, \bibinfo {author} {\bibfnamefont {E.}~\bibnamefont
  {Demler}}, \bibinfo {author} {\bibfnamefont {M.}~\bibnamefont {Lukin}},
  \bibinfo {author} {\bibfnamefont {A.}~\bibnamefont {Micheli}}, \bibinfo
  {author} {\bibfnamefont {N.}~\bibnamefont {Prokof'ev}}, \bibinfo {author}
  {\bibfnamefont {G.}~\bibnamefont {Pupillo}}, \ and\ \bibinfo {author}
  {\bibfnamefont {P.}~\bibnamefont {Zoller}},\ }\href@noop {} {\bibfield
  {journal} {\bibinfo  {journal} {Phys. Rev. Lett.}\ }\textbf {\bibinfo
  {volume} {98}},\ \bibinfo {pages} {60404} (\bibinfo {year}
  {2007})}\BibitemShut {NoStop}%
\bibitem [{\citenamefont {Micheli}\ \emph {et~al.}(2006)\citenamefont
  {Micheli}, \citenamefont {Brennen},\ and\ \citenamefont
  {Zoller}}]{Micheli2006}%
  \BibitemOpen
  \bibfield  {author} {\bibinfo {author} {\bibfnamefont {A.}~\bibnamefont
  {Micheli}}, \bibinfo {author} {\bibfnamefont {G.~K.}\ \bibnamefont
  {Brennen}}, \ and\ \bibinfo {author} {\bibfnamefont {P.}~\bibnamefont
  {Zoller}},\ }\href@noop {} {\bibfield  {journal} {\bibinfo  {journal} {Nat.
  Phys.}\ }\textbf {\bibinfo {volume} {2}},\ \bibinfo {pages} {341} (\bibinfo
  {year} {2006})}\BibitemShut {NoStop}%
\bibitem [{\citenamefont {Ospelkaus}\ \emph {et~al.}(2010)\citenamefont
  {Ospelkaus}, \citenamefont {Ni}, \citenamefont {Wang}, \citenamefont
  {de~Miranda}, \citenamefont {Neyenhuis}, \citenamefont
  {Qu{\'{e}}m{\'{e}}ner}, \citenamefont {Julienne}, \citenamefont {Bohn},
  \citenamefont {Jin},\ and\ \citenamefont {Ye}}]{Ospelkaus2010}%
  \BibitemOpen
  \bibfield  {author} {\bibinfo {author} {\bibfnamefont {S.}~\bibnamefont
  {Ospelkaus}}, \bibinfo {author} {\bibfnamefont {K.-K.}\ \bibnamefont {Ni}},
  \bibinfo {author} {\bibfnamefont {D.}~\bibnamefont {Wang}}, \bibinfo {author}
  {\bibfnamefont {M.~H.~G.}\ \bibnamefont {de~Miranda}}, \bibinfo {author}
  {\bibfnamefont {B.}~\bibnamefont {Neyenhuis}}, \bibinfo {author}
  {\bibfnamefont {G.}~\bibnamefont {Qu{\'{e}}m{\'{e}}ner}}, \bibinfo {author}
  {\bibfnamefont {P.~S.}\ \bibnamefont {Julienne}}, \bibinfo {author}
  {\bibfnamefont {J.~L.}\ \bibnamefont {Bohn}}, \bibinfo {author}
  {\bibfnamefont {D.~S.}\ \bibnamefont {Jin}}, \ and\ \bibinfo {author}
  {\bibfnamefont {J.}~\bibnamefont {Ye}},\ }\href {\doibase
  10.1126/science.1184121} {\bibfield  {journal} {\bibinfo  {journal}
  {Science}\ }\textbf {\bibinfo {volume} {327}},\ \bibinfo {pages} {853}
  (\bibinfo {year} {2010})}\BibitemShut {NoStop}%
\bibitem [{\citenamefont {McDonald}\ \emph {et~al.}(2016)\citenamefont
  {McDonald}, \citenamefont {McGuyer}, \citenamefont {Apfelbeck}, \citenamefont
  {Lee}, \citenamefont {Majewska}, \citenamefont {Moszynski},\ and\
  \citenamefont {Zelevinsky}}]{McDonald2016}%
  \BibitemOpen
  \bibfield  {author} {\bibinfo {author} {\bibfnamefont {M.}~\bibnamefont
  {McDonald}}, \bibinfo {author} {\bibfnamefont {B.~H.}\ \bibnamefont
  {McGuyer}}, \bibinfo {author} {\bibfnamefont {F.}~\bibnamefont {Apfelbeck}},
  \bibinfo {author} {\bibfnamefont {C.-H.}\ \bibnamefont {Lee}}, \bibinfo
  {author} {\bibfnamefont {I.}~\bibnamefont {Majewska}}, \bibinfo {author}
  {\bibfnamefont {R.}~\bibnamefont {Moszynski}}, \ and\ \bibinfo {author}
  {\bibfnamefont {T.}~\bibnamefont {Zelevinsky}},\ }\href@noop {} {\bibfield
  {journal} {\bibinfo  {journal} {Nature}\ }\textbf {\bibinfo {volume} {535}},\
  \bibinfo {pages} {122} (\bibinfo {year} {2016})}\BibitemShut {NoStop}%
\bibitem [{\citenamefont {Hudson}\ \emph {et~al.}(2011)\citenamefont {Hudson},
  \citenamefont {Kara}, \citenamefont {Smallman}, \citenamefont {Sauer},
  \citenamefont {Tarbutt},\ and\ \citenamefont {Hinds}}]{Hudson2011}%
  \BibitemOpen
  \bibfield  {author} {\bibinfo {author} {\bibfnamefont {J.~J.}\ \bibnamefont
  {Hudson}}, \bibinfo {author} {\bibfnamefont {D.~M.}\ \bibnamefont {Kara}},
  \bibinfo {author} {\bibfnamefont {I.}~\bibnamefont {Smallman}}, \bibinfo
  {author} {\bibfnamefont {B.~E.}\ \bibnamefont {Sauer}}, \bibinfo {author}
  {\bibfnamefont {M.~R.}\ \bibnamefont {Tarbutt}}, \ and\ \bibinfo {author}
  {\bibfnamefont {E.~A.}\ \bibnamefont {Hinds}},\ }\href@noop {} {\bibfield
  {journal} {\bibinfo  {journal} {Nature}\ }\textbf {\bibinfo {volume} {473}},\
  \bibinfo {pages} {493} (\bibinfo {year} {2011})}\BibitemShut {NoStop}%
\bibitem [{\citenamefont {Truppe}\ \emph {et~al.}(2013)\citenamefont {Truppe},
  \citenamefont {Hendricks}, \citenamefont {Tokunaga}, \citenamefont
  {Lewandowski}, \citenamefont {Kozlov}, \citenamefont {Henkel}, \citenamefont
  {Hinds},\ and\ \citenamefont {Tarbutt}}]{Truppe2013}%
  \BibitemOpen
  \bibfield  {author} {\bibinfo {author} {\bibfnamefont {S.}~\bibnamefont
  {Truppe}}, \bibinfo {author} {\bibfnamefont {R.}~\bibnamefont {Hendricks}},
  \bibinfo {author} {\bibfnamefont {S.}~\bibnamefont {Tokunaga}}, \bibinfo
  {author} {\bibfnamefont {H.}~\bibnamefont {Lewandowski}}, \bibinfo {author}
  {\bibfnamefont {M.}~\bibnamefont {Kozlov}}, \bibinfo {author} {\bibfnamefont
  {C.}~\bibnamefont {Henkel}}, \bibinfo {author} {\bibfnamefont
  {E.}~\bibnamefont {Hinds}}, \ and\ \bibinfo {author} {\bibfnamefont
  {M.}~\bibnamefont {Tarbutt}},\ }\href@noop {} {\bibfield  {journal} {\bibinfo
   {journal} {Nat. Commun.}\ }\textbf {\bibinfo {volume} {4}} (\bibinfo {year}
  {2013})}\BibitemShut {NoStop}%
\bibitem [{\citenamefont {Hudson}\ \emph {et~al.}(2006)\citenamefont {Hudson},
  \citenamefont {Lewandowski}, \citenamefont {Sawyer},\ and\ \citenamefont
  {Ye}}]{Hudson2006}%
  \BibitemOpen
  \bibfield  {author} {\bibinfo {author} {\bibfnamefont {E.~R.}\ \bibnamefont
  {Hudson}}, \bibinfo {author} {\bibfnamefont {H.~J.}\ \bibnamefont
  {Lewandowski}}, \bibinfo {author} {\bibfnamefont {B.~C.}\ \bibnamefont
  {Sawyer}}, \ and\ \bibinfo {author} {\bibfnamefont {J.}~\bibnamefont {Ye}},\
  }\href@noop {} {\bibfield  {journal} {\bibinfo  {journal} {Phys. Rev. Lett.}\
  }\textbf {\bibinfo {volume} {96}},\ \bibinfo {pages} {143004} (\bibinfo
  {year} {2006})}\BibitemShut {NoStop}%
\bibitem [{\citenamefont {Chin}\ \emph {et~al.}(2010)\citenamefont {Chin},
  \citenamefont {Grimm}, \citenamefont {Julienne},\ and\ \citenamefont
  {Tiesinga}}]{Chin2010}%
  \BibitemOpen
  \bibfield  {author} {\bibinfo {author} {\bibfnamefont {C.}~\bibnamefont
  {Chin}}, \bibinfo {author} {\bibfnamefont {R.}~\bibnamefont {Grimm}},
  \bibinfo {author} {\bibfnamefont {P.}~\bibnamefont {Julienne}}, \ and\
  \bibinfo {author} {\bibfnamefont {E.}~\bibnamefont {Tiesinga}},\ }\href
  {\doibase 10.1103/RevModPhys.82.1225} {\bibfield  {journal} {\bibinfo
  {journal} {Rev. Mod. Phys.}\ }\textbf {\bibinfo {volume} {82}},\ \bibinfo
  {pages} {1225} (\bibinfo {year} {2010})}\BibitemShut {NoStop}%
\bibitem [{\citenamefont {Bergmann}\ \emph {et~al.}(1998)\citenamefont
  {Bergmann}, \citenamefont {Theuer},\ and\ \citenamefont
  {Shore}}]{Bergmann1998}%
  \BibitemOpen
  \bibfield  {author} {\bibinfo {author} {\bibfnamefont {K.}~\bibnamefont
  {Bergmann}}, \bibinfo {author} {\bibfnamefont {H.}~\bibnamefont {Theuer}}, \
  and\ \bibinfo {author} {\bibfnamefont {B.~W.}\ \bibnamefont {Shore}},\ }\href
  {\doibase 10.1103/RevModPhys.70.1003} {\bibfield  {journal} {\bibinfo
  {journal} {Reviews of Modern Physics}\ }\textbf {\bibinfo {volume} {70}},\
  \bibinfo {pages} {1003} (\bibinfo {year} {1998})}\BibitemShut {NoStop}%
\bibitem [{\citenamefont {Sage}\ \emph {et~al.}(2005)\citenamefont {Sage},
  \citenamefont {Sainis}, \citenamefont {Bergeman},\ and\ \citenamefont
  {DeMille}}]{Sage2005}%
  \BibitemOpen
  \bibfield  {author} {\bibinfo {author} {\bibfnamefont {J.~M.}\ \bibnamefont
  {Sage}}, \bibinfo {author} {\bibfnamefont {S.}~\bibnamefont {Sainis}},
  \bibinfo {author} {\bibfnamefont {T.}~\bibnamefont {Bergeman}}, \ and\
  \bibinfo {author} {\bibfnamefont {D.}~\bibnamefont {DeMille}},\ }\href@noop
  {} {\bibfield  {journal} {\bibinfo  {journal} {Phys. Rev. Lett.}\ }\textbf
  {\bibinfo {volume} {94}},\ \bibinfo {pages} {203001} (\bibinfo {year}
  {2005})}\BibitemShut {NoStop}%
\bibitem [{\citenamefont {Ni}\ \emph {et~al.}(2008)\citenamefont {Ni},
  \citenamefont {Ospelkaus}, \citenamefont {de~Miranda}, \citenamefont {Pe'er},
  \citenamefont {Neyenhuis}, \citenamefont {Zirbel}, \citenamefont
  {Kotochigova}, \citenamefont {Julienne}, \citenamefont {Jin},\ and\
  \citenamefont {Ye}}]{Ni2008}%
  \BibitemOpen
  \bibfield  {author} {\bibinfo {author} {\bibfnamefont {K.-K.}\ \bibnamefont
  {Ni}}, \bibinfo {author} {\bibfnamefont {S.}~\bibnamefont {Ospelkaus}},
  \bibinfo {author} {\bibfnamefont {M.~H.~G.}\ \bibnamefont {de~Miranda}},
  \bibinfo {author} {\bibfnamefont {A.}~\bibnamefont {Pe'er}}, \bibinfo
  {author} {\bibfnamefont {B.}~\bibnamefont {Neyenhuis}}, \bibinfo {author}
  {\bibfnamefont {J.~J.}\ \bibnamefont {Zirbel}}, \bibinfo {author}
  {\bibfnamefont {S.}~\bibnamefont {Kotochigova}}, \bibinfo {author}
  {\bibfnamefont {P.~S.}\ \bibnamefont {Julienne}}, \bibinfo {author}
  {\bibfnamefont {D.~S.}\ \bibnamefont {Jin}}, \ and\ \bibinfo {author}
  {\bibfnamefont {J.}~\bibnamefont {Ye}},\ }\href@noop {} {\bibfield  {journal}
  {\bibinfo  {journal} {Science}\ }\textbf {\bibinfo {volume} {322}},\ \bibinfo
  {pages} {231} (\bibinfo {year} {2008})}\BibitemShut {NoStop}%
\bibitem [{\citenamefont {Danzl}\ \emph {et~al.}(2010)\citenamefont {Danzl},
  \citenamefont {Mark}, \citenamefont {Haller}, \citenamefont {Gustavsson},
  \citenamefont {Hart}, \citenamefont {Aldegunde}, \citenamefont {Hutson},\
  and\ \citenamefont {N\"agerl}}]{Danzl2010}%
  \BibitemOpen
  \bibfield  {author} {\bibinfo {author} {\bibfnamefont {J.~G.}\ \bibnamefont
  {Danzl}}, \bibinfo {author} {\bibfnamefont {M.~J.}\ \bibnamefont {Mark}},
  \bibinfo {author} {\bibfnamefont {E.}~\bibnamefont {Haller}}, \bibinfo
  {author} {\bibfnamefont {M.}~\bibnamefont {Gustavsson}}, \bibinfo {author}
  {\bibfnamefont {R.}~\bibnamefont {Hart}}, \bibinfo {author} {\bibfnamefont
  {J.}~\bibnamefont {Aldegunde}}, \bibinfo {author} {\bibfnamefont {J.~M.}\
  \bibnamefont {Hutson}}, \ and\ \bibinfo {author} {\bibfnamefont {H.-C.}\
  \bibnamefont {N\"agerl}},\ }\href {\doibase doi:10.1038/nphys1533} {\bibfield
   {journal} {\bibinfo  {journal} {Nat. Phys.}\ }\textbf {\bibinfo {volume}
  {6}},\ \bibinfo {pages} {265} (\bibinfo {year} {2010})}\BibitemShut {NoStop}%
\bibitem [{\citenamefont {Wu}\ \emph {et~al.}(2014)\citenamefont {Wu},
  \citenamefont {Park}, \citenamefont {Ahmadi}, \citenamefont {Will},\ and\
  \citenamefont {Zwierlein}}]{Wu2014}%
  \BibitemOpen
  \bibfield  {author} {\bibinfo {author} {\bibfnamefont {C.-H.}\ \bibnamefont
  {Wu}}, \bibinfo {author} {\bibfnamefont {J.~W.}\ \bibnamefont {Park}},
  \bibinfo {author} {\bibfnamefont {P.}~\bibnamefont {Ahmadi}}, \bibinfo
  {author} {\bibfnamefont {S.}~\bibnamefont {Will}}, \ and\ \bibinfo {author}
  {\bibfnamefont {M.~W.}\ \bibnamefont {Zwierlein}},\ }\href {\doibase
  10.1103/PhysRevLett.109.085301} {\bibfield  {journal} {\bibinfo  {journal}
  {Phys. Rev. Lett.}\ }\textbf {\bibinfo {volume} {109}},\ \bibinfo {pages}
  {85301} (\bibinfo {year} {2014})}\BibitemShut {NoStop}%
\bibitem [{\citenamefont {Molony}\ \emph {et~al.}(2014)\citenamefont {Molony},
  \citenamefont {Gregory}, \citenamefont {Ji}, \citenamefont {Lu},
  \citenamefont {K{\"{o}}ppinger}, \citenamefont {{Le Sueur}}, \citenamefont
  {Blackley}, \citenamefont {Hutson},\ and\ \citenamefont
  {Cornish}}]{Molony2014}%
  \BibitemOpen
  \bibfield  {author} {\bibinfo {author} {\bibfnamefont {P.~K.}\ \bibnamefont
  {Molony}}, \bibinfo {author} {\bibfnamefont {P.~D.}\ \bibnamefont {Gregory}},
  \bibinfo {author} {\bibfnamefont {Z.}~\bibnamefont {Ji}}, \bibinfo {author}
  {\bibfnamefont {B.}~\bibnamefont {Lu}}, \bibinfo {author} {\bibfnamefont
  {M.~P.}\ \bibnamefont {K{\"{o}}ppinger}}, \bibinfo {author} {\bibfnamefont
  {C.~R.}\ \bibnamefont {{Le Sueur}}}, \bibinfo {author} {\bibfnamefont
  {C.~L.}\ \bibnamefont {Blackley}}, \bibinfo {author} {\bibfnamefont {J.~M.}\
  \bibnamefont {Hutson}}, \ and\ \bibinfo {author} {\bibfnamefont {S.~L.}\
  \bibnamefont {Cornish}},\ }\href {\doibase 10.1103/PhysRevLett.113.255301}
  {\bibfield  {journal} {\bibinfo  {journal} {Physical review letters}\
  }\textbf {\bibinfo {volume} {113}},\ \bibinfo {pages} {255301} (\bibinfo
  {year} {2014})},\ \Eprint {http://arxiv.org/abs/1409.1485} {arXiv:1409.1485}
  \BibitemShut {NoStop}%
\bibitem [{\citenamefont {Takekoshi}\ \emph {et~al.}(2014)\citenamefont
  {Takekoshi}, \citenamefont {Reichs{\"o}llner}, \citenamefont {Schindewolf},
  \citenamefont {Hutson}, \citenamefont {Le~Sueur}, \citenamefont {Dulieu},
  \citenamefont {Ferlaino}, \citenamefont {Grimm},\ and\ \citenamefont
  {N{\"a}gerl}}]{Takekoshi2014}%
  \BibitemOpen
  \bibfield  {author} {\bibinfo {author} {\bibfnamefont {T.}~\bibnamefont
  {Takekoshi}}, \bibinfo {author} {\bibfnamefont {L.}~\bibnamefont
  {Reichs{\"o}llner}}, \bibinfo {author} {\bibfnamefont {A.}~\bibnamefont
  {Schindewolf}}, \bibinfo {author} {\bibfnamefont {J.~M.}\ \bibnamefont
  {Hutson}}, \bibinfo {author} {\bibfnamefont {C.~R.}\ \bibnamefont
  {Le~Sueur}}, \bibinfo {author} {\bibfnamefont {O.}~\bibnamefont {Dulieu}},
  \bibinfo {author} {\bibfnamefont {F.}~\bibnamefont {Ferlaino}}, \bibinfo
  {author} {\bibfnamefont {R.}~\bibnamefont {Grimm}}, \ and\ \bibinfo {author}
  {\bibfnamefont {H.-C.}\ \bibnamefont {N{\"a}gerl}},\ }\href@noop {}
  {\bibfield  {journal} {\bibinfo  {journal} {Phys. Rev. Lett.}\ }\textbf
  {\bibinfo {volume} {113}},\ \bibinfo {pages} {205301} (\bibinfo {year}
  {2014})}\BibitemShut {NoStop}%
\bibitem [{\citenamefont {Guo}\ \emph {et~al.}(2016)\citenamefont {Guo},
  \citenamefont {Zhu}, \citenamefont {Lu}, \citenamefont {Ye}, \citenamefont
  {Wang}, \citenamefont {Vexiau}, \citenamefont {Bouloufa-Maafa}, \citenamefont
  {Qu{\'{e}}m{\'{e}}ner}, \citenamefont {Dulieu},\ and\ \citenamefont
  {Wang}}]{Guo2016}%
  \BibitemOpen
  \bibfield  {author} {\bibinfo {author} {\bibfnamefont {M.}~\bibnamefont
  {Guo}}, \bibinfo {author} {\bibfnamefont {B.}~\bibnamefont {Zhu}}, \bibinfo
  {author} {\bibfnamefont {B.}~\bibnamefont {Lu}}, \bibinfo {author}
  {\bibfnamefont {X.}~\bibnamefont {Ye}}, \bibinfo {author} {\bibfnamefont
  {F.}~\bibnamefont {Wang}}, \bibinfo {author} {\bibfnamefont {R.}~\bibnamefont
  {Vexiau}}, \bibinfo {author} {\bibfnamefont {N.}~\bibnamefont
  {Bouloufa-Maafa}}, \bibinfo {author} {\bibfnamefont {G.}~\bibnamefont
  {Qu{\'{e}}m{\'{e}}ner}}, \bibinfo {author} {\bibfnamefont {O.}~\bibnamefont
  {Dulieu}}, \ and\ \bibinfo {author} {\bibfnamefont {D.}~\bibnamefont
  {Wang}},\ }\href {\doibase 10.1103/PhysRevLett.116.205303} {\bibfield
  {journal} {\bibinfo  {journal} {Phys. Rev. Lett.}\ }\textbf {\bibinfo
  {volume} {116}},\ \bibinfo {pages} {205303} (\bibinfo {year}
  {2016})}\BibitemShut {NoStop}%
\bibitem [{\citenamefont {Zuchowski}\ \emph {et~al.}(2010)\citenamefont
  {Zuchowski}, \citenamefont {Aldegunde},\ and\ \citenamefont
  {Hutson}}]{Zuchowski2010}%
  \BibitemOpen
  \bibfield  {author} {\bibinfo {author} {\bibfnamefont {P.~S.}\ \bibnamefont
  {Zuchowski}}, \bibinfo {author} {\bibfnamefont {J.}~\bibnamefont
  {Aldegunde}}, \ and\ \bibinfo {author} {\bibfnamefont {J.}~\bibnamefont
  {Hutson}},\ }\href {\doibase 10.1103/PhysRevLett.105.153201} {\bibfield
  {journal} {\bibinfo  {journal} {Phys. Rev. Lett.}\ }\textbf {\bibinfo
  {volume} {105}} (\bibinfo {year} {2010}),\
  10.1103/PhysRevLett.105.153201}\BibitemShut {NoStop}%
\bibitem [{\citenamefont {Brue}\ and\ \citenamefont {Hutson}(2012)}]{Brue2012}%
  \BibitemOpen
  \bibfield  {author} {\bibinfo {author} {\bibfnamefont {D.~A.}\ \bibnamefont
  {Brue}}\ and\ \bibinfo {author} {\bibfnamefont {J.~M.}\ \bibnamefont
  {Hutson}},\ }\href {\doibase 10.1103/PhysRevLett.108.043201} {\bibfield
  {journal} {\bibinfo  {journal} {Phys. Rev. Lett.}\ }\textbf {\bibinfo
  {volume} {108}},\ \bibinfo {pages} {43201} (\bibinfo {year}
  {2012})}\BibitemShut {NoStop}%
\bibitem [{\citenamefont {Tomza}\ \emph {et~al.}(2014)\citenamefont {Tomza},
  \citenamefont {Gonz{\'{a}}lez-F{\'{e}}rez}, \citenamefont {Koch},\ and\
  \citenamefont {Moszynski}}]{Tomza2014}%
  \BibitemOpen
  \bibfield  {author} {\bibinfo {author} {\bibfnamefont {M.}~\bibnamefont
  {Tomza}}, \bibinfo {author} {\bibfnamefont {R.}~\bibnamefont
  {Gonz{\'{a}}lez-F{\'{e}}rez}}, \bibinfo {author} {\bibfnamefont {C.~P.}\
  \bibnamefont {Koch}}, \ and\ \bibinfo {author} {\bibfnamefont
  {R.}~\bibnamefont {Moszynski}},\ }\href {\doibase
  10.1103/PhysRevLett.112.113201} {\bibfield  {journal} {\bibinfo  {journal}
  {Phys. Rev. Lett.}\ }\textbf {\bibinfo {volume} {112}},\ \bibinfo {pages}
  {113201} (\bibinfo {year} {2014})}\BibitemShut {NoStop}%
\bibitem [{\citenamefont {Tassy}\ \emph {et~al.}(2010)\citenamefont {Tassy},
  \citenamefont {Nemitz}, \citenamefont {Baumer}, \citenamefont {H{\"{o}}hl},
  \citenamefont {Bat{\"{a}}r},\ and\ \citenamefont
  {G{\"{o}}rlitz}}]{Tassy2010}%
  \BibitemOpen
  \bibfield  {author} {\bibinfo {author} {\bibfnamefont {S.}~\bibnamefont
  {Tassy}}, \bibinfo {author} {\bibfnamefont {N.}~\bibnamefont {Nemitz}},
  \bibinfo {author} {\bibfnamefont {F.}~\bibnamefont {Baumer}}, \bibinfo
  {author} {\bibfnamefont {C.}~\bibnamefont {H{\"{o}}hl}}, \bibinfo {author}
  {\bibfnamefont {A.}~\bibnamefont {Bat{\"{a}}r}}, \ and\ \bibinfo {author}
  {\bibfnamefont {A.}~\bibnamefont {G{\"{o}}rlitz}},\ }\href
  {http://stacks.iop.org/0953-4075/43/i=20/a=205309} {\bibfield  {journal}
  {\bibinfo  {journal} {J. Phys. B}\ }\textbf {\bibinfo {volume} {43}},\
  \bibinfo {pages} {205309} (\bibinfo {year} {2010})}\BibitemShut {NoStop}%
\bibitem [{\citenamefont {Baumer}\ \emph {et~al.}(2011)\citenamefont {Baumer},
  \citenamefont {M{\"{u}}nchow}, \citenamefont {G{\"{o}}rlitz}, \citenamefont
  {Maxwell}, \citenamefont {Julienne},\ and\ \citenamefont
  {Tiesinga}}]{Baumer2011}%
  \BibitemOpen
  \bibfield  {author} {\bibinfo {author} {\bibfnamefont {F.}~\bibnamefont
  {Baumer}}, \bibinfo {author} {\bibfnamefont {F.}~\bibnamefont
  {M{\"{u}}nchow}}, \bibinfo {author} {\bibfnamefont {A.}~\bibnamefont
  {G{\"{o}}rlitz}}, \bibinfo {author} {\bibfnamefont {S.~E.}\ \bibnamefont
  {Maxwell}}, \bibinfo {author} {\bibfnamefont {P.~S.}\ \bibnamefont
  {Julienne}}, \ and\ \bibinfo {author} {\bibfnamefont {E.}~\bibnamefont
  {Tiesinga}},\ }\href {\doibase 10.1103/PhysRevA.83.040702} {\bibfield
  {journal} {\bibinfo  {journal} {Phys. Rev. A}\ }\textbf {\bibinfo {volume}
  {83}},\ \bibinfo {pages} {40702} (\bibinfo {year} {2011})}\BibitemShut
  {NoStop}%
\bibitem [{\citenamefont {Nemitz}\ \emph {et~al.}(2009)\citenamefont {Nemitz},
  \citenamefont {Baumer}, \citenamefont {M{\"{u}}nchow}, \citenamefont
  {Tassy},\ and\ \citenamefont {G{\"{o}}rlitz}}]{Nemitz2009}%
  \BibitemOpen
  \bibfield  {author} {\bibinfo {author} {\bibfnamefont {N.}~\bibnamefont
  {Nemitz}}, \bibinfo {author} {\bibfnamefont {F.}~\bibnamefont {Baumer}},
  \bibinfo {author} {\bibfnamefont {F.}~\bibnamefont {M{\"{u}}nchow}}, \bibinfo
  {author} {\bibfnamefont {S.}~\bibnamefont {Tassy}}, \ and\ \bibinfo {author}
  {\bibfnamefont {A.}~\bibnamefont {G{\"{o}}rlitz}},\ }\href {\doibase
  10.1103/PhysRevA.79.061403} {\bibfield  {journal} {\bibinfo  {journal} {Phys.
  Rev. A}\ }\textbf {\bibinfo {volume} {79}},\ \bibinfo {pages} {61403}
  (\bibinfo {year} {2009})}\BibitemShut {NoStop}%
\bibitem [{\citenamefont {Munchow}\ \emph {et~al.}(2011)\citenamefont
  {Munchow}, \citenamefont {Bruni}, \citenamefont {Madalinski},\ and\
  \citenamefont {Gorlitz}}]{Munchow2011}%
  \BibitemOpen
  \bibfield  {author} {\bibinfo {author} {\bibfnamefont {F.}~\bibnamefont
  {Munchow}}, \bibinfo {author} {\bibfnamefont {C.}~\bibnamefont {Bruni}},
  \bibinfo {author} {\bibfnamefont {M.}~\bibnamefont {Madalinski}}, \ and\
  \bibinfo {author} {\bibfnamefont {A.}~\bibnamefont {Gorlitz}},\ }\href
  {\doibase 10.1039/C1CP21219B} {\bibfield  {journal} {\bibinfo  {journal}
  {Phys. Chem. Chem. Phys.}\ }\textbf {\bibinfo {volume} {13}},\ \bibinfo
  {pages} {18734} (\bibinfo {year} {2011})}\BibitemShut {NoStop}%
\bibitem [{\citenamefont {Borkowski}\ \emph {et~al.}(2013)\citenamefont
  {Borkowski}, \citenamefont {{\.{Z}}uchowski}, \citenamefont {Ciury{\l}o},
  \citenamefont {Julienne}, \citenamefont {K{\c{e}}dziera}, \citenamefont
  {Mentel}, \citenamefont {Tecmer}, \citenamefont {M{\"{u}}nchow},
  \citenamefont {Bruni},\ and\ \citenamefont {G{\"{o}}rlitz}}]{Borkowski2013}%
  \BibitemOpen
  \bibfield  {author} {\bibinfo {author} {\bibfnamefont {M.}~\bibnamefont
  {Borkowski}}, \bibinfo {author} {\bibfnamefont {P.~S.}\ \bibnamefont
  {{\.{Z}}uchowski}}, \bibinfo {author} {\bibfnamefont {R.}~\bibnamefont
  {Ciury{\l}o}}, \bibinfo {author} {\bibfnamefont {P.~S.}\ \bibnamefont
  {Julienne}}, \bibinfo {author} {\bibfnamefont {D.}~\bibnamefont
  {K{\c{e}}dziera}}, \bibinfo {author} {\bibfnamefont {{\L}.}~\bibnamefont
  {Mentel}}, \bibinfo {author} {\bibfnamefont {P.}~\bibnamefont {Tecmer}},
  \bibinfo {author} {\bibfnamefont {F.}~\bibnamefont {M{\"{u}}nchow}}, \bibinfo
  {author} {\bibfnamefont {C.}~\bibnamefont {Bruni}}, \ and\ \bibinfo {author}
  {\bibfnamefont {A.}~\bibnamefont {G{\"{o}}rlitz}},\ }\href {\doibase
  10.1103/PhysRevA.88.052708} {\bibfield  {journal} {\bibinfo  {journal} {Phys.
  Rev. A}\ }\textbf {\bibinfo {volume} {88}},\ \bibinfo {pages} {052708}
  (\bibinfo {year} {2013})}\BibitemShut {NoStop}%
\bibitem [{\citenamefont {Pasquiou}\ \emph {et~al.}(2013)\citenamefont
  {Pasquiou}, \citenamefont {Bayerle}, \citenamefont {Tzanova}, \citenamefont
  {Stellmer}, \citenamefont {Szczepkowski}, \citenamefont {Parigger},
  \citenamefont {Grimm},\ and\ \citenamefont {Schreck}}]{Pasquiou2013}%
  \BibitemOpen
  \bibfield  {author} {\bibinfo {author} {\bibfnamefont {B.}~\bibnamefont
  {Pasquiou}}, \bibinfo {author} {\bibfnamefont {A.}~\bibnamefont {Bayerle}},
  \bibinfo {author} {\bibfnamefont {S.~M.}\ \bibnamefont {Tzanova}}, \bibinfo
  {author} {\bibfnamefont {S.}~\bibnamefont {Stellmer}}, \bibinfo {author}
  {\bibfnamefont {J.}~\bibnamefont {Szczepkowski}}, \bibinfo {author}
  {\bibfnamefont {M.}~\bibnamefont {Parigger}}, \bibinfo {author}
  {\bibfnamefont {R.}~\bibnamefont {Grimm}}, \ and\ \bibinfo {author}
  {\bibfnamefont {F.}~\bibnamefont {Schreck}},\ }\href {\doibase
  10.1103/PhysRevA.88.023601} {\bibfield  {journal} {\bibinfo  {journal} {Phys.
  Rev. A}\ }\textbf {\bibinfo {volume} {88}},\ \bibinfo {pages} {023601}
  (\bibinfo {year} {2013})}\BibitemShut {NoStop}%
\bibitem [{\citenamefont {Kemp}\ \emph {et~al.}(2016)\citenamefont {Kemp},
  \citenamefont {Butler}, \citenamefont {Freytag}, \citenamefont {Hopkins},
  \citenamefont {Hinds}, \citenamefont {Tarbutt},\ and\ \citenamefont
  {Cornish}}]{Kemp2016}%
  \BibitemOpen
  \bibfield  {author} {\bibinfo {author} {\bibfnamefont {S.}~\bibnamefont
  {Kemp}}, \bibinfo {author} {\bibfnamefont {K.}~\bibnamefont {Butler}},
  \bibinfo {author} {\bibfnamefont {R.}~\bibnamefont {Freytag}}, \bibinfo
  {author} {\bibfnamefont {S.}~\bibnamefont {Hopkins}}, \bibinfo {author}
  {\bibfnamefont {E.}~\bibnamefont {Hinds}}, \bibinfo {author} {\bibfnamefont
  {M.}~\bibnamefont {Tarbutt}}, \ and\ \bibinfo {author} {\bibfnamefont
  {S.}~\bibnamefont {Cornish}},\ }\href@noop {} {\bibfield  {journal} {\bibinfo
   {journal} {Rev. Sci. Inst.}\ }\textbf {\bibinfo {volume} {87}},\ \bibinfo
  {pages} {023105} (\bibinfo {year} {2016})}\BibitemShut {NoStop}%
\bibitem [{\citenamefont {Guttridge}\ \emph {et~al.}(2017)\citenamefont
  {Guttridge}, \citenamefont {Hopkins}, \citenamefont {Kemp}, \citenamefont
  {Frye}, \citenamefont {Hutson},\ and\ \citenamefont
  {Cornish}}]{Guttridge2017}%
  \BibitemOpen
  \bibfield  {author} {\bibinfo {author} {\bibfnamefont {A.}~\bibnamefont
  {Guttridge}}, \bibinfo {author} {\bibfnamefont {S.}~\bibnamefont {Hopkins}},
  \bibinfo {author} {\bibfnamefont {S.}~\bibnamefont {Kemp}}, \bibinfo {author}
  {\bibfnamefont {M.~D.}\ \bibnamefont {Frye}}, \bibinfo {author}
  {\bibfnamefont {J.~M.}\ \bibnamefont {Hutson}}, \ and\ \bibinfo {author}
  {\bibfnamefont {S.~L.}\ \bibnamefont {Cornish}},\ }\href@noop {} {\bibfield
  {journal} {\bibinfo  {journal} {arXiv preprint arXiv:1704.03270}\ } (\bibinfo
  {year} {2017})}\BibitemShut {NoStop}%
\bibitem [{\citenamefont {Hara}\ \emph {et~al.}(2011)\citenamefont {Hara},
  \citenamefont {Takasu}, \citenamefont {Yamaoka}, \citenamefont {Doyle},\ and\
  \citenamefont {Takahashi}}]{Hara2011}%
  \BibitemOpen
  \bibfield  {author} {\bibinfo {author} {\bibfnamefont {H.}~\bibnamefont
  {Hara}}, \bibinfo {author} {\bibfnamefont {Y.}~\bibnamefont {Takasu}},
  \bibinfo {author} {\bibfnamefont {Y.}~\bibnamefont {Yamaoka}}, \bibinfo
  {author} {\bibfnamefont {J.~M.}\ \bibnamefont {Doyle}}, \ and\ \bibinfo
  {author} {\bibfnamefont {Y.}~\bibnamefont {Takahashi}},\ }\href {\doibase
  10.1103/PhysRevLett.106.205304} {\bibfield  {journal} {\bibinfo  {journal}
  {Phys. Rev. Lett.}\ }\textbf {\bibinfo {volume} {106}},\ \bibinfo {pages}
  {205304} (\bibinfo {year} {2011})}\BibitemShut {NoStop}%
\bibitem [{\citenamefont {Hansen}\ \emph {et~al.}(2013)\citenamefont {Hansen},
  \citenamefont {Khramov}, \citenamefont {Dowd}, \citenamefont {Jamison},
  \citenamefont {Plotkin-Swing}, \citenamefont {Roy},\ and\ \citenamefont
  {Gupta}}]{Hansen2013}%
  \BibitemOpen
  \bibfield  {author} {\bibinfo {author} {\bibfnamefont {A.~H.}\ \bibnamefont
  {Hansen}}, \bibinfo {author} {\bibfnamefont {A.~Y.}\ \bibnamefont {Khramov}},
  \bibinfo {author} {\bibfnamefont {W.~H.}\ \bibnamefont {Dowd}}, \bibinfo
  {author} {\bibfnamefont {A.~O.}\ \bibnamefont {Jamison}}, \bibinfo {author}
  {\bibfnamefont {B.}~\bibnamefont {Plotkin-Swing}}, \bibinfo {author}
  {\bibfnamefont {R.~J.}\ \bibnamefont {Roy}}, \ and\ \bibinfo {author}
  {\bibfnamefont {S.}~\bibnamefont {Gupta}},\ }\href {\doibase
  10.1103/PhysRevA.87.013615} {\bibfield  {journal} {\bibinfo  {journal} {Phys.
  Rev. A}\ }\textbf {\bibinfo {volume} {87}},\ \bibinfo {pages} {013615}
  (\bibinfo {year} {2013})}\BibitemShut {NoStop}%
\bibitem [{\citenamefont {Roy}\ \emph {et~al.}(2016)\citenamefont {Roy},
  \citenamefont {Shrestha}, \citenamefont {Green}, \citenamefont {Gupta},
  \citenamefont {Li}, \citenamefont {Kotochigova}, \citenamefont {Petrov},\
  and\ \citenamefont {Yuen}}]{Roy2016}%
  \BibitemOpen
  \bibfield  {author} {\bibinfo {author} {\bibfnamefont {R.}~\bibnamefont
  {Roy}}, \bibinfo {author} {\bibfnamefont {R.}~\bibnamefont {Shrestha}},
  \bibinfo {author} {\bibfnamefont {A.}~\bibnamefont {Green}}, \bibinfo
  {author} {\bibfnamefont {S.}~\bibnamefont {Gupta}}, \bibinfo {author}
  {\bibfnamefont {M.}~\bibnamefont {Li}}, \bibinfo {author} {\bibfnamefont
  {S.}~\bibnamefont {Kotochigova}}, \bibinfo {author} {\bibfnamefont
  {A.}~\bibnamefont {Petrov}}, \ and\ \bibinfo {author} {\bibfnamefont {C.~H.}\
  \bibnamefont {Yuen}},\ }\href {\doibase 10.1103/PhysRevA.94.033413}
  {\bibfield  {journal} {\bibinfo  {journal} {Phys. Rev. A}\ }\textbf {\bibinfo
  {volume} {94}},\ \bibinfo {pages} {033413} (\bibinfo {year}
  {2016})}\BibitemShut {NoStop}%
\bibitem [{\citenamefont {Witkowski}\ \emph {et~al.}(2017)\citenamefont
  {Witkowski}, \citenamefont {Nag{\'o}rny}, \citenamefont {Munoz-Rodriguez},
  \citenamefont {Ciury{\l}o}, \citenamefont {{\.Z}uchowski}, \citenamefont
  {Bilicki}, \citenamefont {Piotrowski}, \citenamefont {Morzy{\'n}ski},\ and\
  \citenamefont {Zawada}}]{Witkowski2017}%
  \BibitemOpen
  \bibfield  {author} {\bibinfo {author} {\bibfnamefont {M.}~\bibnamefont
  {Witkowski}}, \bibinfo {author} {\bibfnamefont {B.}~\bibnamefont
  {Nag{\'o}rny}}, \bibinfo {author} {\bibfnamefont {R.}~\bibnamefont
  {Munoz-Rodriguez}}, \bibinfo {author} {\bibfnamefont {R.}~\bibnamefont
  {Ciury{\l}o}}, \bibinfo {author} {\bibfnamefont {P.~S.}\ \bibnamefont
  {{\.Z}uchowski}}, \bibinfo {author} {\bibfnamefont {S.}~\bibnamefont
  {Bilicki}}, \bibinfo {author} {\bibfnamefont {M.}~\bibnamefont {Piotrowski}},
  \bibinfo {author} {\bibfnamefont {P.}~\bibnamefont {Morzy{\'n}ski}}, \ and\
  \bibinfo {author} {\bibfnamefont {M.}~\bibnamefont {Zawada}},\ }\href@noop {}
  {\bibfield  {journal} {\bibinfo  {journal} {Opt. Express}\ }\textbf {\bibinfo
  {volume} {25}},\ \bibinfo {pages} {3165} (\bibinfo {year}
  {2017})}\BibitemShut {NoStop}%
\bibitem [{\citenamefont {Meyer}\ and\ \citenamefont {Bohn}(2009)}]{Meyer2009}%
  \BibitemOpen
  \bibfield  {author} {\bibinfo {author} {\bibfnamefont {E.~R.}\ \bibnamefont
  {Meyer}}\ and\ \bibinfo {author} {\bibfnamefont {J.~L.}\ \bibnamefont
  {Bohn}},\ }\href {\doibase 10.1103/PhysRevA.80.042508} {\bibfield  {journal}
  {\bibinfo  {journal} {Physical Review A}\ }\textbf {\bibinfo {volume} {80}},\
  \bibinfo {pages} {042508} (\bibinfo {year} {2009})}\BibitemShut {NoStop}%
\bibitem [{\citenamefont {Kro{\'{s}}nicki}\ \emph {et~al.}(2015)\citenamefont
  {Kro{\'{s}}nicki}, \citenamefont {Strojecki}, \citenamefont
  {Urba{\'{n}}czyk}, \citenamefont {Pashov},\ and\ \citenamefont
  {Koperski}}]{Krosnicki2015}%
  \BibitemOpen
  \bibfield  {author} {\bibinfo {author} {\bibfnamefont {M.}~\bibnamefont
  {Kro{\'{s}}nicki}}, \bibinfo {author} {\bibfnamefont {M.}~\bibnamefont
  {Strojecki}}, \bibinfo {author} {\bibfnamefont {T.}~\bibnamefont
  {Urba{\'{n}}czyk}}, \bibinfo {author} {\bibfnamefont {A.}~\bibnamefont
  {Pashov}}, \ and\ \bibinfo {author} {\bibfnamefont {J.}~\bibnamefont
  {Koperski}},\ }\href {\doibase 10.1016/j.physrep.2015.06.004} {\bibfield
  {journal} {\bibinfo  {journal} {Physics Reports}\ }\textbf {\bibinfo {volume}
  {591}},\ \bibinfo {pages} {1} (\bibinfo {year} {2015})}\BibitemShut {NoStop}%
\bibitem [{\citenamefont {Marvet}\ and\ \citenamefont
  {Dantus}(1995)}]{Marvet1995}%
  \BibitemOpen
  \bibfield  {author} {\bibinfo {author} {\bibfnamefont {U.}~\bibnamefont
  {Marvet}}\ and\ \bibinfo {author} {\bibfnamefont {M.}~\bibnamefont
  {Dantus}},\ }\href
  {https://ac.els-cdn.com/0009261495010185/1-s2.0-0009261495010185-main.pdf?{\_}tid=11825bb6-a6a7-11e7-ba1a-00000aab0f27{\&}acdnat=1506862394{\_}2cd221483b91a710a803881674eca3b7}
  {\bibfield  {journal} {\bibinfo  {journal} {Chem. Phys. Lett.}\ }\textbf
  {\bibinfo {volume} {245}},\ \bibinfo {pages} {393} (\bibinfo {year}
  {1995})}\BibitemShut {NoStop}%
\bibitem [{\citenamefont {Gaston}\ and\ \citenamefont
  {Schwerdtfeger}(2006)}]{Gaston2006}%
  \BibitemOpen
  \bibfield  {author} {\bibinfo {author} {\bibfnamefont {N.}~\bibnamefont
  {Gaston}}\ and\ \bibinfo {author} {\bibfnamefont {P.}~\bibnamefont
  {Schwerdtfeger}},\ }\href@noop {} {\bibfield  {journal} {\bibinfo  {journal}
  {Phys. Rev. B}\ }\textbf {\bibinfo {volume} {74}},\ \bibinfo {pages} {024105}
  (\bibinfo {year} {2006})}\BibitemShut {NoStop}%
\bibitem [{\citenamefont {Figgen}\ \emph {et~al.}(2005)\citenamefont {Figgen},
  \citenamefont {Rauhut}, \citenamefont {Dolg},\ and\ \citenamefont
  {Stoll}}]{Figgen2005}%
  \BibitemOpen
  \bibfield  {author} {\bibinfo {author} {\bibfnamefont {D.}~\bibnamefont
  {Figgen}}, \bibinfo {author} {\bibfnamefont {G.}~\bibnamefont {Rauhut}},
  \bibinfo {author} {\bibfnamefont {M.}~\bibnamefont {Dolg}}, \ and\ \bibinfo
  {author} {\bibfnamefont {H.}~\bibnamefont {Stoll}},\ }\href@noop {}
  {\bibfield  {journal} {\bibinfo  {journal} {J. Chem. Phys.}\ }\textbf
  {\bibinfo {volume} {122}},\ \bibinfo {pages} {104103} (\bibinfo {year}
  {2005})}\BibitemShut {NoStop}%
\bibitem [{\citenamefont {Peterson}\ and\ \citenamefont
  {Puzzarini}(2005)}]{Peterson:05}%
  \BibitemOpen
  \bibfield  {author} {\bibinfo {author} {\bibfnamefont {K.}~\bibnamefont
  {Peterson}}\ and\ \bibinfo {author} {\bibfnamefont {C.}~\bibnamefont
  {Puzzarini}},\ }\href@noop {} {\bibfield  {journal} {\bibinfo  {journal}
  {Theor. Chem. Acc.}\ }\textbf {\bibinfo {volume} {144}},\ \bibinfo {pages}
  {283} (\bibinfo {year} {2005})}\BibitemShut {NoStop}%
\bibitem [{\citenamefont {Gruber}\ \emph {et~al.}(1994)\citenamefont {Gruber},
  \citenamefont {Musso}, \citenamefont {Windholz}, \citenamefont {Gleichmann},
  \citenamefont {Hess}, \citenamefont {Fuso},\ and\ \citenamefont
  {Allegrini}}]{Gruber1994}%
  \BibitemOpen
  \bibfield  {author} {\bibinfo {author} {\bibfnamefont {D.}~\bibnamefont
  {Gruber}}, \bibinfo {author} {\bibfnamefont {M.}~\bibnamefont {Musso}},
  \bibinfo {author} {\bibfnamefont {L.}~\bibnamefont {Windholz}}, \bibinfo
  {author} {\bibfnamefont {M.}~\bibnamefont {Gleichmann}}, \bibinfo {author}
  {\bibfnamefont {B.}~\bibnamefont {Hess}}, \bibinfo {author} {\bibfnamefont
  {F.}~\bibnamefont {Fuso}}, \ and\ \bibinfo {author} {\bibfnamefont
  {M.}~\bibnamefont {Allegrini}},\ }\href@noop {} {\bibfield  {journal}
  {\bibinfo  {journal} {J. Chem. Phys.}\ }\textbf {\bibinfo {volume} {101}},\
  \bibinfo {pages} {929} (\bibinfo {year} {1994})}\BibitemShut {NoStop}%
\bibitem [{\citenamefont {Gruber}\ \emph {et~al.}(1996)\citenamefont {Gruber},
  \citenamefont {Li}, \citenamefont {Windholz}, \citenamefont {Gleichmann},
  \citenamefont {Hess}, \citenamefont {Vezmar},\ and\ \citenamefont
  {Pichler}}]{Gruber1996}%
  \BibitemOpen
  \bibfield  {author} {\bibinfo {author} {\bibfnamefont {D.}~\bibnamefont
  {Gruber}}, \bibinfo {author} {\bibfnamefont {X.}~\bibnamefont {Li}}, \bibinfo
  {author} {\bibfnamefont {L.}~\bibnamefont {Windholz}}, \bibinfo {author}
  {\bibfnamefont {M.}~\bibnamefont {Gleichmann}}, \bibinfo {author}
  {\bibfnamefont {B.}~\bibnamefont {Hess}}, \bibinfo {author} {\bibfnamefont
  {I.}~\bibnamefont {Vezmar}}, \ and\ \bibinfo {author} {\bibfnamefont
  {G.}~\bibnamefont {Pichler}},\ }\href@noop {} {\bibfield  {journal} {\bibinfo
   {journal} {J. Chem. Phys.}\ }\textbf {\bibinfo {volume} {100}},\ \bibinfo
  {pages} {10062} (\bibinfo {year} {1996})}\BibitemShut {NoStop}%
\bibitem [{\citenamefont {Polly}\ \emph {et~al.}(1998)\citenamefont {Polly},
  \citenamefont {Gruber}, \citenamefont {Windholz}, \citenamefont
  {Gleichmann},\ and\ \citenamefont {He{\ss}}}]{Polly1998}%
  \BibitemOpen
  \bibfield  {author} {\bibinfo {author} {\bibfnamefont {R.}~\bibnamefont
  {Polly}}, \bibinfo {author} {\bibfnamefont {D.}~\bibnamefont {Gruber}},
  \bibinfo {author} {\bibfnamefont {L.}~\bibnamefont {Windholz}}, \bibinfo
  {author} {\bibfnamefont {M.~M.}\ \bibnamefont {Gleichmann}}, \ and\ \bibinfo
  {author} {\bibfnamefont {B.~A.}\ \bibnamefont {He{\ss}}},\ }\href@noop {}
  {\bibfield  {journal} {\bibinfo  {journal} {J. Chem. Phys.}\ }\textbf
  {\bibinfo {volume} {109}},\ \bibinfo {pages} {9463} (\bibinfo {year}
  {1998})}\BibitemShut {NoStop}%
\bibitem [{\citenamefont {Fedichev}\ \emph {et~al.}(1996)\citenamefont
  {Fedichev}, \citenamefont {Kagan}, \citenamefont {Shlyapnikov},\ and\
  \citenamefont {Walraven}}]{Fedichev1996}%
  \BibitemOpen
  \bibfield  {author} {\bibinfo {author} {\bibfnamefont {P.~O.}\ \bibnamefont
  {Fedichev}}, \bibinfo {author} {\bibfnamefont {Y.}~\bibnamefont {Kagan}},
  \bibinfo {author} {\bibfnamefont {G.~V.}\ \bibnamefont {Shlyapnikov}}, \ and\
  \bibinfo {author} {\bibfnamefont {J.~T.~M.}\ \bibnamefont {Walraven}},\
  }\href {\doibase 10.1103/PhysRevLett.77.2913} {\bibfield  {journal} {\bibinfo
   {journal} {Physical Review Letters}\ }\textbf {\bibinfo {volume} {77}},\
  \bibinfo {pages} {12} (\bibinfo {year} {1996})},\ \Eprint
  {http://arxiv.org/abs/9605008} {arXiv:9605008 [atom-ph]} \BibitemShut
  {NoStop}%
\bibitem [{\citenamefont {Bohn}\ and\ \citenamefont
  {Julienne}(1997)}]{Bohn1997}%
  \BibitemOpen
  \bibfield  {author} {\bibinfo {author} {\bibfnamefont {J.}~\bibnamefont
  {Bohn}}\ and\ \bibinfo {author} {\bibfnamefont {P.}~\bibnamefont
  {Julienne}},\ }\href {\doibase 10.1103/PhysRevA.56.1486} {\bibfield
  {journal} {\bibinfo  {journal} {Physical Review A}\ }\textbf {\bibinfo
  {volume} {56}},\ \bibinfo {pages} {1486} (\bibinfo {year}
  {1997})}\BibitemShut {NoStop}%
\bibitem [{\citenamefont {Ciury{\l}o}\ \emph {et~al.}(2005)\citenamefont
  {Ciury{\l}o}, \citenamefont {Tiesinga},\ and\ \citenamefont
  {Julienne}}]{Ciurylo2005}%
  \BibitemOpen
  \bibfield  {author} {\bibinfo {author} {\bibfnamefont {R.}~\bibnamefont
  {Ciury{\l}o}}, \bibinfo {author} {\bibfnamefont {E.}~\bibnamefont
  {Tiesinga}}, \ and\ \bibinfo {author} {\bibfnamefont {P.~S.}\ \bibnamefont
  {Julienne}},\ }\href {\doibase 10.1103/PhysRevA.71.030701} {\bibfield
  {journal} {\bibinfo  {journal} {Physical Review A - Atomic, Molecular, and
  Optical Physics}\ }\textbf {\bibinfo {volume} {71}},\ \bibinfo {pages}
  {030701(R)} (\bibinfo {year} {2005})},\ \Eprint
  {http://arxiv.org/abs/0412111} {arXiv:0412111 [physics]} \BibitemShut
  {NoStop}%
\bibitem [{\citenamefont {Kokoouline}\ \emph {et~al.}(2001)\citenamefont
  {Kokoouline}, \citenamefont {Vala},\ and\ \citenamefont
  {Kosloff}}]{Kokoouline2001}%
  \BibitemOpen
  \bibfield  {author} {\bibinfo {author} {\bibfnamefont {V.}~\bibnamefont
  {Kokoouline}}, \bibinfo {author} {\bibfnamefont {J.}~\bibnamefont {Vala}}, \
  and\ \bibinfo {author} {\bibfnamefont {R.}~\bibnamefont {Kosloff}},\ }\href
  {\doibase 10.1063/1.1343080} {\bibfield  {journal} {\bibinfo  {journal} {J.
  Chem. Phys.}\ }\textbf {\bibinfo {volume} {114}},\ \bibinfo {pages} {3046}
  (\bibinfo {year} {2001})}\BibitemShut {NoStop}%
\bibitem [{\citenamefont {Knowles}\ \emph {et~al.}(1993)\citenamefont
  {Knowles}, \citenamefont {Hampel},\ and\ \citenamefont
  {Werner}}]{Knowles1993}%
  \BibitemOpen
  \bibfield  {author} {\bibinfo {author} {\bibfnamefont {P.~J.}\ \bibnamefont
  {Knowles}}, \bibinfo {author} {\bibfnamefont {C.}~\bibnamefont {Hampel}}, \
  and\ \bibinfo {author} {\bibfnamefont {H.~J.}\ \bibnamefont {Werner}},\
  }\href@noop {} {\bibfield  {journal} {\bibinfo  {journal} {J. Chem. Phys.}\
  }\textbf {\bibinfo {volume} {99}},\ \bibinfo {pages} {5219} (\bibinfo {year}
  {1993})}\BibitemShut {NoStop}%
\bibitem [{\citenamefont {Werner}\ \emph {et~al.}(2012)\citenamefont {Werner},
  \citenamefont {Knowles} \emph {et~al.}}]{MOLPRO_brief:2012}%
  \BibitemOpen
  \bibfield  {author} {\bibinfo {author} {\bibfnamefont {H.-J.}\ \bibnamefont
  {Werner}}, \bibinfo {author} {\bibfnamefont {P.~J.}\ \bibnamefont {Knowles}},
   \emph {et~al.},\ }\href@noop {} {\enquote {\bibinfo {title} {{\sc MOLPRO},
  version 2012.1.12: A package of ab initio programs},}\ } (\bibinfo {year}
  {2012}),\ \bibinfo {note} {see http://www.molpro.net}\BibitemShut {NoStop}%
\bibitem [{\citenamefont {Stanton}\ and\ \citenamefont
  {Bartlett}(1993)}]{Stanton1993}%
  \BibitemOpen
  \bibfield  {author} {\bibinfo {author} {\bibfnamefont {J.~F.}\ \bibnamefont
  {Stanton}}\ and\ \bibinfo {author} {\bibfnamefont {R.~J.}\ \bibnamefont
  {Bartlett}},\ }\href {\doibase 10.1063/1.464746} {\bibfield  {journal}
  {\bibinfo  {journal} {J. Chem. Phys.}\ }\textbf {\bibinfo {volume} {98}},\
  \bibinfo {pages} {7029} (\bibinfo {year} {1993})}\BibitemShut {NoStop}%
\bibitem [{\citenamefont {Bussery-Honvault}\ and\ \citenamefont
  {Moszynski}(2006)}]{Bussery-Honvault2006}%
  \BibitemOpen
  \bibfield  {author} {\bibinfo {author} {\bibfnamefont {B.}~\bibnamefont
  {Bussery-Honvault}}\ and\ \bibinfo {author} {\bibfnamefont {R.}~\bibnamefont
  {Moszynski}},\ }\href {\doibase 10.1080/00268970600674023} {\bibfield
  {journal} {\bibinfo  {journal} {Mol. Phys.}\ }\textbf {\bibinfo {volume}
  {104}},\ \bibinfo {pages} {2387} (\bibinfo {year} {2006})}\BibitemShut
  {NoStop}%
\bibitem [{\citenamefont {Skomorowski}\ \emph {et~al.}(2012)\citenamefont
  {Skomorowski}, \citenamefont {Paw{\l}owski}, \citenamefont {Koch},\ and\
  \citenamefont {Moszynski}}]{Skomorowski2012}%
  \BibitemOpen
  \bibfield  {author} {\bibinfo {author} {\bibfnamefont {W.}~\bibnamefont
  {Skomorowski}}, \bibinfo {author} {\bibfnamefont {F.}~\bibnamefont
  {Paw{\l}owski}}, \bibinfo {author} {\bibfnamefont {C.~P.}\ \bibnamefont
  {Koch}}, \ and\ \bibinfo {author} {\bibfnamefont {R.}~\bibnamefont
  {Moszynski}},\ }\href {\doibase 10.1063/1.4713939} {\bibfield  {journal}
  {\bibinfo  {journal} {The Journal of Chemical Physics}\ }\textbf {\bibinfo
  {volume} {136}},\ \bibinfo {pages} {194306} (\bibinfo {year}
  {2012})}\BibitemShut {NoStop}%
\bibitem [{\citenamefont {Pototschnig}\ \emph {et~al.}(2017)\citenamefont
  {Pototschnig}, \citenamefont {Meyer}, \citenamefont {Hauser},\ and\
  \citenamefont {Ernst}}]{Pototschnig2017}%
  \BibitemOpen
  \bibfield  {author} {\bibinfo {author} {\bibfnamefont {J.~V.}\ \bibnamefont
  {Pototschnig}}, \bibinfo {author} {\bibfnamefont {R.}~\bibnamefont {Meyer}},
  \bibinfo {author} {\bibfnamefont {A.~W.}\ \bibnamefont {Hauser}}, \ and\
  \bibinfo {author} {\bibfnamefont {W.~E.}\ \bibnamefont {Ernst}},\ }\href@noop
  {} {\bibfield  {journal} {\bibinfo  {journal} {Phys. Rev. A}\ }\textbf
  {\bibinfo {volume} {95}},\ \bibinfo {pages} {022501} (\bibinfo {year}
  {2017})}\BibitemShut {NoStop}%
\bibitem [{\citenamefont {Boys}\ and\ \citenamefont
  {Bernardi}(1970)}]{Boys:1970}%
  \BibitemOpen
  \bibfield  {author} {\bibinfo {author} {\bibfnamefont {S.}~\bibnamefont
  {Boys}}\ and\ \bibinfo {author} {\bibfnamefont {F.}~\bibnamefont
  {Bernardi}},\ }\href@noop {} {\bibfield  {journal} {\bibinfo  {journal} {Mol.
  Phys.}\ }\textbf {\bibinfo {volume} {19}},\ \bibinfo {pages} {553} (\bibinfo
  {year} {1970})}\BibitemShut {NoStop}%
\bibitem [{\citenamefont {Lim}\ \emph {et~al.}(2005)\citenamefont {Lim},
  \citenamefont {Schwerdtfeger}, \citenamefont {Metz},\ and\ \citenamefont
  {Stoll}}]{Lim2005}%
  \BibitemOpen
  \bibfield  {author} {\bibinfo {author} {\bibfnamefont {I.}~\bibnamefont
  {Lim}}, \bibinfo {author} {\bibfnamefont {P.}~\bibnamefont {Schwerdtfeger}},
  \bibinfo {author} {\bibfnamefont {B.}~\bibnamefont {Metz}}, \ and\ \bibinfo
  {author} {\bibfnamefont {H.}~\bibnamefont {Stoll}},\ }\href@noop {}
  {\bibfield  {journal} {\bibinfo  {journal} {J. Chem. Phys.}\ }\textbf
  {\bibinfo {volume} {122}},\ \bibinfo {pages} {104103} (\bibinfo {year}
  {2005})}\BibitemShut {NoStop}%
\bibitem [{\citenamefont {{\.{Z}}uchowski}\ \emph {et~al.}(2014)\citenamefont
  {{\.{Z}}uchowski}, \citenamefont {Gu{\'{e}}rout},\ and\ \citenamefont
  {Dulieu}}]{Zuchowski2014}%
  \BibitemOpen
  \bibfield  {author} {\bibinfo {author} {\bibfnamefont {P.~S.}\ \bibnamefont
  {{\.{Z}}uchowski}}, \bibinfo {author} {\bibfnamefont {R.}~\bibnamefont
  {Gu{\'{e}}rout}}, \ and\ \bibinfo {author} {\bibfnamefont {O.}~\bibnamefont
  {Dulieu}},\ }\href {\doibase 10.1103/PhysRevA.90.012507} {\bibfield
  {journal} {\bibinfo  {journal} {Physical Review A}\ }\textbf {\bibinfo
  {volume} {90}},\ \bibinfo {pages} {012507} (\bibinfo {year}
  {2014})}\BibitemShut {NoStop}%
\bibitem [{\citenamefont {Helgaker}\ \emph {et~al.}(2000)\citenamefont
  {Helgaker}, \citenamefont {Joergensen},\ and\ \citenamefont
  {Olsen}}]{Helgaker:2000}%
  \BibitemOpen
  \bibfield  {author} {\bibinfo {author} {\bibfnamefont {T.}~\bibnamefont
  {Helgaker}}, \bibinfo {author} {\bibfnamefont {P.}~\bibnamefont
  {Joergensen}}, \ and\ \bibinfo {author} {\bibfnamefont {J.}~\bibnamefont
  {Olsen}},\ }\href@noop {} {\emph {\bibinfo {title} {Molecular
  Electronic-Structure Theory}}}\ (\bibinfo  {publisher} {Wiley},\ \bibinfo
  {address} {Chichester},\ \bibinfo {year} {2000})\BibitemShut {NoStop}%
\bibitem [{\citenamefont {Koperski}\ \emph {et~al.}(1994)\citenamefont
  {Koperski}, \citenamefont {Atkinson},\ and\ \citenamefont
  {Krause}}]{Koperski:94}%
  \BibitemOpen
  \bibfield  {author} {\bibinfo {author} {\bibfnamefont {J.}~\bibnamefont
  {Koperski}}, \bibinfo {author} {\bibfnamefont {J.}~\bibnamefont {Atkinson}},
  \ and\ \bibinfo {author} {\bibfnamefont {L.}~\bibnamefont {Krause}},\
  }\href@noop {} {\bibfield  {journal} {\bibinfo  {journal} {Chem. Phys.
  Lett.}\ }\textbf {\bibinfo {volume} {161}},\ \bibinfo {pages} {219} (\bibinfo
  {year} {1994})}\BibitemShut {NoStop}%
\bibitem [{\citenamefont {Koperski}\ \emph {et~al.}(1997)\citenamefont
  {Koperski}, \citenamefont {Atkinson},\ and\ \citenamefont
  {Krause}}]{Koperski:97}%
  \BibitemOpen
  \bibfield  {author} {\bibinfo {author} {\bibfnamefont {J.}~\bibnamefont
  {Koperski}}, \bibinfo {author} {\bibfnamefont {J.}~\bibnamefont {Atkinson}},
  \ and\ \bibinfo {author} {\bibfnamefont {L.}~\bibnamefont {Krause}},\
  }\href@noop {} {\bibfield  {journal} {\bibinfo  {journal} {J. Mol.
  Spectrosc.}\ }\textbf {\bibinfo {volume} {184}},\ \bibinfo {pages} {300}
  (\bibinfo {year} {1997})}\BibitemShut {NoStop}%
\bibitem [{\citenamefont {Koperski}\ \emph {et~al.}(2008)\citenamefont
  {Koperski}, \citenamefont {Qu}, \citenamefont {Meng}, \citenamefont
  {Kenefick},\ and\ \citenamefont {E.}}]{Koperski:08}%
  \BibitemOpen
  \bibfield  {author} {\bibinfo {author} {\bibfnamefont {J.}~\bibnamefont
  {Koperski}}, \bibinfo {author} {\bibfnamefont {X.}~\bibnamefont {Qu}},
  \bibinfo {author} {\bibfnamefont {H.}~\bibnamefont {Meng}}, \bibinfo {author}
  {\bibfnamefont {R.}~\bibnamefont {Kenefick}}, \ and\ \bibinfo {author}
  {\bibfnamefont {F.}~\bibnamefont {E.}},\ }\href@noop {} {\bibfield  {journal}
  {\bibinfo  {journal} {Chem. Phys.}\ }\textbf {\bibinfo {volume} {348}},\
  \bibinfo {pages} {103} (\bibinfo {year} {2008})}\BibitemShut {NoStop}%
\bibitem [{\citenamefont {Pahl}\ \emph {et~al.}(2011)\citenamefont {Pahl},
  \citenamefont {Figgen}, \citenamefont {Borschevsky}, \citenamefont
  {Peterson},\ and\ \citenamefont {Schwerdtfeger}}]{Figgen:11}%
  \BibitemOpen
  \bibfield  {author} {\bibinfo {author} {\bibfnamefont {E.}~\bibnamefont
  {Pahl}}, \bibinfo {author} {\bibfnamefont {D.}~\bibnamefont {Figgen}},
  \bibinfo {author} {\bibfnamefont {A.}~\bibnamefont {Borschevsky}}, \bibinfo
  {author} {\bibfnamefont {K.}~\bibnamefont {Peterson}}, \ and\ \bibinfo
  {author} {\bibfnamefont {P.}~\bibnamefont {Schwerdtfeger}},\ }\href@noop {}
  {\bibfield  {journal} {\bibinfo  {journal} {Theor. Chem. Acc.}\ }\textbf
  {\bibinfo {volume} {129}},\ \bibinfo {pages} {651} (\bibinfo {year}
  {2011})}\BibitemShut {NoStop}%
\bibitem [{\citenamefont {Stanton}\ \emph {et~al.}()\citenamefont {Stanton},
  \citenamefont {Gauss}, \citenamefont {Harding}, \citenamefont {Szalay} \emph
  {et~al.}}]{CFOUR_brief}%
  \BibitemOpen
  \bibfield  {author} {\bibinfo {author} {\bibfnamefont {J.}~\bibnamefont
  {Stanton}}, \bibinfo {author} {\bibfnamefont {J.}~\bibnamefont {Gauss}},
  \bibinfo {author} {\bibfnamefont {M.}~\bibnamefont {Harding}}, \bibinfo
  {author} {\bibfnamefont {P.}~\bibnamefont {Szalay}},  \emph {et~al.},\
  }\href@noop {} {\enquote {\bibinfo {title} {Cfour, a quantum chemical program
  package, http://www.cfour.de},}\ }\BibitemShut {NoStop}%
\bibitem [{\citenamefont {S{\o}rensen}\ \emph {et~al.}(2009)\citenamefont
  {S{\o}rensen}, \citenamefont {Knecht}, \citenamefont {Fleig},\ and\
  \citenamefont {Marian}}]{Sorensen2009}%
  \BibitemOpen
  \bibfield  {author} {\bibinfo {author} {\bibfnamefont {L.~K.}\ \bibnamefont
  {S{\o}rensen}}, \bibinfo {author} {\bibfnamefont {S.}~\bibnamefont {Knecht}},
  \bibinfo {author} {\bibfnamefont {T.}~\bibnamefont {Fleig}}, \ and\ \bibinfo
  {author} {\bibfnamefont {C.~M.}\ \bibnamefont {Marian}},\ }\href@noop {}
  {\bibfield  {journal} {\bibinfo  {journal} {J. Phys. Chem. A}\ }\textbf
  {\bibinfo {volume} {113}},\ \bibinfo {pages} {12607} (\bibinfo {year}
  {2009})}\BibitemShut {NoStop}%
\bibitem [{\citenamefont {Derevianko}\ \emph {et~al.}(2010)\citenamefont
  {Derevianko}, \citenamefont {Porsev},\ and\ \citenamefont
  {Babb}}]{Derevianko:2010}%
  \BibitemOpen
  \bibfield  {author} {\bibinfo {author} {\bibfnamefont {A.}~\bibnamefont
  {Derevianko}}, \bibinfo {author} {\bibfnamefont {S.~G.}\ \bibnamefont
  {Porsev}}, \ and\ \bibinfo {author} {\bibfnamefont {J.~F.}\ \bibnamefont
  {Babb}},\ }\href@noop {} {\bibfield  {journal} {\bibinfo  {journal} {At. Data
  Nucl. Data Tables}\ }\textbf {\bibinfo {volume} {96}} (\bibinfo {year}
  {2010})}\BibitemShut {NoStop}%
\bibitem [{\citenamefont {Moszynski}\ \emph {et~al.}(2005)\citenamefont
  {Moszynski}, \citenamefont {\.Zuchowski},\ and\ \citenamefont
  {Jeziorski}}]{Moszynski:2005}%
  \BibitemOpen
  \bibfield  {author} {\bibinfo {author} {\bibfnamefont {R.}~\bibnamefont
  {Moszynski}}, \bibinfo {author} {\bibfnamefont {P.~S.}\ \bibnamefont
  {\.Zuchowski}}, \ and\ \bibinfo {author} {\bibfnamefont {B.}~\bibnamefont
  {Jeziorski}},\ }\href@noop {} {\bibfield  {journal} {\bibinfo  {journal}
  {Collect. Czech. Chem. Commun.}\ }\textbf {\bibinfo {volume} {70}},\ \bibinfo
  {pages} {1109} (\bibinfo {year} {2005})}\BibitemShut {NoStop}%
\bibitem [{\citenamefont {Korona}\ \emph {et~al.}(2006)\citenamefont {Korona},
  \citenamefont {Przybytek},\ and\ \citenamefont {Jeziorski}}]{Korona:06}%
  \BibitemOpen
  \bibfield  {author} {\bibinfo {author} {\bibfnamefont {T.}~\bibnamefont
  {Korona}}, \bibinfo {author} {\bibfnamefont {M.}~\bibnamefont {Przybytek}}, \
  and\ \bibinfo {author} {\bibfnamefont {B.}~\bibnamefont {Jeziorski}},\
  }\href@noop {} {\bibfield  {journal} {\bibinfo  {journal} {Mol. Phys.}\
  }\textbf {\bibinfo {volume} {104}},\ \bibinfo {pages} {2303} (\bibinfo {year}
  {2006})}\BibitemShut {NoStop}%
\bibitem [{\citenamefont {Jiang}\ \emph {et~al.}(2013)\citenamefont {Jiang},
  \citenamefont {Cheng},\ and\ \citenamefont {Mitroy}}]{Jiang2013}%
  \BibitemOpen
  \bibfield  {author} {\bibinfo {author} {\bibfnamefont {J.}~\bibnamefont
  {Jiang}}, \bibinfo {author} {\bibfnamefont {Y.}~\bibnamefont {Cheng}}, \ and\
  \bibinfo {author} {\bibfnamefont {J.}~\bibnamefont {Mitroy}},\ }\href@noop {}
  {\bibfield  {journal} {\bibinfo  {journal} {J. Phys. B}\ }\textbf {\bibinfo
  {volume} {46}},\ \bibinfo {pages} {125004} (\bibinfo {year}
  {2013})}\BibitemShut {NoStop}%
\bibitem [{\citenamefont {Pople}\ \emph {et~al.}(1968)\citenamefont {Pople},
  \citenamefont {McIver},\ and\ \citenamefont {Ostlund}}]{Pople1968}%
  \BibitemOpen
  \bibfield  {author} {\bibinfo {author} {\bibfnamefont {J.~A.}\ \bibnamefont
  {Pople}}, \bibinfo {author} {\bibfnamefont {J.~W.}\ \bibnamefont {McIver}}, \
  and\ \bibinfo {author} {\bibfnamefont {N.~S.}\ \bibnamefont {Ostlund}},\
  }\href {\doibase 10.1063/1.1670536} {\bibfield  {journal} {\bibinfo
  {journal} {J. Chem. Phys.}\ }\textbf {\bibinfo {volume} {49}},\ \bibinfo
  {pages} {2960} (\bibinfo {year} {1968})}\BibitemShut {NoStop}%
\bibitem [{\citenamefont {{\.{Z}}uchowski}\ \emph {et~al.}(2013)\citenamefont
  {{\.{Z}}uchowski}, \citenamefont {Kosicki}, \citenamefont {Kodrycka},\ and\
  \citenamefont {Sold{\'{a}}n}}]{Zuchowski2013}%
  \BibitemOpen
  \bibfield  {author} {\bibinfo {author} {\bibfnamefont {P.~S.}\ \bibnamefont
  {{\.{Z}}uchowski}}, \bibinfo {author} {\bibfnamefont {M.}~\bibnamefont
  {Kosicki}}, \bibinfo {author} {\bibfnamefont {M.}~\bibnamefont {Kodrycka}}, \
  and\ \bibinfo {author} {\bibfnamefont {P.}~\bibnamefont {Sold{\'{a}}n}},\
  }\href {\doibase 10.1103/PhysRevA.87.022706} {\bibfield  {journal} {\bibinfo
  {journal} {Phys. Rev. A}\ }\textbf {\bibinfo {volume} {87}},\ \bibinfo
  {pages} {022706} (\bibinfo {year} {2013})}\BibitemShut {NoStop}%
\bibitem [{\citenamefont {Ciury{\l}o}\ \emph {et~al.}(2004)\citenamefont
  {Ciury{\l}o}, \citenamefont {Tiesinga}, \citenamefont {Kotochigova},\ and\
  \citenamefont {Julienne}}]{Ciurylo2004}%
  \BibitemOpen
  \bibfield  {author} {\bibinfo {author} {\bibfnamefont {R.}~\bibnamefont
  {Ciury{\l}o}}, \bibinfo {author} {\bibfnamefont {E.}~\bibnamefont
  {Tiesinga}}, \bibinfo {author} {\bibfnamefont {S.}~\bibnamefont
  {Kotochigova}}, \ and\ \bibinfo {author} {\bibfnamefont {P.}~\bibnamefont
  {Julienne}},\ }\href {\doibase 10.1103/PhysRevA.70.062710} {\bibfield
  {journal} {\bibinfo  {journal} {Physical Review A}\ }\textbf {\bibinfo
  {volume} {70}},\ \bibinfo {pages} {062710} (\bibinfo {year}
  {2004})}\BibitemShut {NoStop}%
\bibitem [{\citenamefont {Mies}(1973)}]{Mies1973}%
  \BibitemOpen
  \bibfield  {author} {\bibinfo {author} {\bibfnamefont {F.~H.}\ \bibnamefont
  {Mies}},\ }\href {\doibase 10.1103/PhysRevA.7.942} {\bibfield  {journal}
  {\bibinfo  {journal} {Physical Review A}\ }\textbf {\bibinfo {volume} {7}},\
  \bibinfo {pages} {942} (\bibinfo {year} {1973})}\BibitemShut {NoStop}%
\bibitem [{\citenamefont {Gao}(1996)}]{Gao1996}%
  \BibitemOpen
  \bibfield  {author} {\bibinfo {author} {\bibfnamefont {B.}~\bibnamefont
  {Gao}},\ }\href {\doibase 10.1103/PhysRevA.54.2022} {\bibfield  {journal}
  {\bibinfo  {journal} {Physical Review A}\ }\textbf {\bibinfo {volume} {54}},\
  \bibinfo {pages} {2022} (\bibinfo {year} {1996})}\BibitemShut {NoStop}%
\bibitem [{\citenamefont {Borkowski}\ \emph {et~al.}(2009)\citenamefont
  {Borkowski}, \citenamefont {Ciury{\l}o}, \citenamefont {Julienne},
  \citenamefont {Tojo}, \citenamefont {Enomoto},\ and\ \citenamefont
  {Takahashi}}]{Borkowski2009}%
  \BibitemOpen
  \bibfield  {author} {\bibinfo {author} {\bibfnamefont {M.}~\bibnamefont
  {Borkowski}}, \bibinfo {author} {\bibfnamefont {R.}~\bibnamefont
  {Ciury{\l}o}}, \bibinfo {author} {\bibfnamefont {P.~S.}\ \bibnamefont
  {Julienne}}, \bibinfo {author} {\bibfnamefont {S.}~\bibnamefont {Tojo}},
  \bibinfo {author} {\bibfnamefont {K.}~\bibnamefont {Enomoto}}, \ and\
  \bibinfo {author} {\bibfnamefont {Y.}~\bibnamefont {Takahashi}},\ }\href
  {\doibase 10.1103/PhysRevA.80.012715} {\bibfield  {journal} {\bibinfo
  {journal} {Physical Review A - Atomic, Molecular, and Optical Physics}\
  }\textbf {\bibinfo {volume} {80}},\ \bibinfo {pages} {012715} (\bibinfo
  {year} {2009})},\ \Eprint {http://arxiv.org/abs/0905.0958} {arXiv:0905.0958}
  \BibitemShut {NoStop}%
\bibitem [{\citenamefont {Gribakin}\ and\ \citenamefont
  {Flambaum}(1993)}]{Gribakin1993}%
  \BibitemOpen
  \bibfield  {author} {\bibinfo {author} {\bibfnamefont {G.~F.}\ \bibnamefont
  {Gribakin}}\ and\ \bibinfo {author} {\bibfnamefont {V.~V.}\ \bibnamefont
  {Flambaum}},\ }\href {\doibase 10.1103/PhysRevA.48.546} {\bibfield  {journal}
  {\bibinfo  {journal} {Physical Review A}\ }\textbf {\bibinfo {volume} {48}},\
  \bibinfo {pages} {546} (\bibinfo {year} {1993})}\BibitemShut {NoStop}%
\bibitem [{\citenamefont {Kitagawa}\ \emph {et~al.}(2008)\citenamefont
  {Kitagawa}, \citenamefont {Enomoto}, \citenamefont {Kasa}, \citenamefont
  {Takahashi}, \citenamefont {Ciury{\l}o}, \citenamefont {Naidon},\ and\
  \citenamefont {Julienne}}]{Kitagawa2008}%
  \BibitemOpen
  \bibfield  {author} {\bibinfo {author} {\bibfnamefont {M.}~\bibnamefont
  {Kitagawa}}, \bibinfo {author} {\bibfnamefont {K.}~\bibnamefont {Enomoto}},
  \bibinfo {author} {\bibfnamefont {K.}~\bibnamefont {Kasa}}, \bibinfo {author}
  {\bibfnamefont {Y.}~\bibnamefont {Takahashi}}, \bibinfo {author}
  {\bibfnamefont {R.}~\bibnamefont {Ciury{\l}o}}, \bibinfo {author}
  {\bibfnamefont {P.}~\bibnamefont {Naidon}}, \ and\ \bibinfo {author}
  {\bibfnamefont {P.~S.}\ \bibnamefont {Julienne}},\ }\href {\doibase
  10.1103/PhysRevA.77.012719} {\bibfield  {journal} {\bibinfo  {journal}
  {Physical Review A - Atomic, Molecular, and Optical Physics}\ }\textbf
  {\bibinfo {volume} {77}},\ \bibinfo {pages} {012719} (\bibinfo {year}
  {2008})},\ \Eprint {http://arxiv.org/abs/0708.0752} {arXiv:0708.0752}
  \BibitemShut {NoStop}%
\bibitem [{\citenamefont {Bohn}\ and\ \citenamefont
  {Julienne}(1999)}]{Bohn1999}%
  \BibitemOpen
  \bibfield  {author} {\bibinfo {author} {\bibfnamefont {J.}~\bibnamefont
  {Bohn}}\ and\ \bibinfo {author} {\bibfnamefont {P.}~\bibnamefont
  {Julienne}},\ }\href {\doibase 10.1103/PhysRevA.60.414} {\bibfield  {journal}
  {\bibinfo  {journal} {Physical Review A}\ }\textbf {\bibinfo {volume} {60}},\
  \bibinfo {pages} {414} (\bibinfo {year} {1999})}\BibitemShut {NoStop}%
\bibitem [{\citenamefont {Yan}\ \emph {et~al.}(2013)\citenamefont {Yan},
  \citenamefont {Desalvo}, \citenamefont {Huang}, \citenamefont {Naidon},\ and\
  \citenamefont {Killian}}]{Yan2013}%
  \BibitemOpen
  \bibfield  {author} {\bibinfo {author} {\bibfnamefont {M.}~\bibnamefont
  {Yan}}, \bibinfo {author} {\bibfnamefont {B.~J.}\ \bibnamefont {Desalvo}},
  \bibinfo {author} {\bibfnamefont {Y.}~\bibnamefont {Huang}}, \bibinfo
  {author} {\bibfnamefont {P.}~\bibnamefont {Naidon}}, \ and\ \bibinfo {author}
  {\bibfnamefont {T.~C.}\ \bibnamefont {Killian}},\ }\href {\doibase
  10.1103/PhysRevLett.111.150402} {\bibfield  {journal} {\bibinfo  {journal}
  {Physical Review Letters}\ }\textbf {\bibinfo {volume} {111}},\ \bibinfo
  {pages} {150402} (\bibinfo {year} {2013})},\ \Eprint
  {http://arxiv.org/abs/1308.4118} {arXiv:1308.4118} \BibitemShut {NoStop}%
\bibitem [{\citenamefont {Jones}\ \emph {et~al.}(2006)\citenamefont {Jones},
  \citenamefont {Tiesinga}, \citenamefont {Lett},\ and\ \citenamefont
  {Julienne}}]{Jones2006}%
  \BibitemOpen
  \bibfield  {author} {\bibinfo {author} {\bibfnamefont {K.~M.}\ \bibnamefont
  {Jones}}, \bibinfo {author} {\bibfnamefont {E.}~\bibnamefont {Tiesinga}},
  \bibinfo {author} {\bibfnamefont {P.~D.}\ \bibnamefont {Lett}}, \ and\
  \bibinfo {author} {\bibfnamefont {P.~S.}\ \bibnamefont {Julienne}},\ }\href
  {\doibase 10.1103/RevModPhys.78.483} {\bibfield  {journal} {\bibinfo
  {journal} {Reviews of Modern Physics}\ }\textbf {\bibinfo {volume} {78}},\
  \bibinfo {pages} {483} (\bibinfo {year} {2006})}\BibitemShut {NoStop}%
\bibitem [{\citenamefont {Nicholson}\ \emph {et~al.}(2015)\citenamefont
  {Nicholson}, \citenamefont {Blatt}, \citenamefont {Bloom}, \citenamefont
  {Williams}, \citenamefont {Thomsen}, \citenamefont {Ye},\ and\ \citenamefont
  {Julienne}}]{Nicholson2015}%
  \BibitemOpen
  \bibfield  {author} {\bibinfo {author} {\bibfnamefont {T.~L.}\ \bibnamefont
  {Nicholson}}, \bibinfo {author} {\bibfnamefont {S.}~\bibnamefont {Blatt}},
  \bibinfo {author} {\bibfnamefont {B.~J.}\ \bibnamefont {Bloom}}, \bibinfo
  {author} {\bibfnamefont {J.~R.}\ \bibnamefont {Williams}}, \bibinfo {author}
  {\bibfnamefont {J.~W.}\ \bibnamefont {Thomsen}}, \bibinfo {author}
  {\bibfnamefont {J.}~\bibnamefont {Ye}}, \ and\ \bibinfo {author}
  {\bibfnamefont {P.~S.}\ \bibnamefont {Julienne}},\ }\href {\doibase
  10.1103/PhysRevA.92.022709} {\bibfield  {journal} {\bibinfo  {journal}
  {Physical Review A - Atomic, Molecular, and Optical Physics}\ }\textbf
  {\bibinfo {volume} {92}},\ \bibinfo {pages} {022709} (\bibinfo {year}
  {2015})},\ \Eprint {http://arxiv.org/abs/1502.00026} {arXiv:1502.00026}
  \BibitemShut {NoStop}%
\bibitem [{\citenamefont {Jones}\ \emph {et~al.}(1999)\citenamefont {Jones},
  \citenamefont {Lett}, \citenamefont {Tiesinga},\ and\ \citenamefont
  {Julienne}}]{Jones1999}%
  \BibitemOpen
  \bibfield  {author} {\bibinfo {author} {\bibfnamefont {K.~M.}\ \bibnamefont
  {Jones}}, \bibinfo {author} {\bibfnamefont {P.~D.}\ \bibnamefont {Lett}},
  \bibinfo {author} {\bibfnamefont {E.}~\bibnamefont {Tiesinga}}, \ and\
  \bibinfo {author} {\bibfnamefont {P.~S.}\ \bibnamefont {Julienne}},\ }\href
  {\doibase 10.1103/PhysRevA.61.012501} {\bibfield  {journal} {\bibinfo
  {journal} {Physical Review A}\ }\textbf {\bibinfo {volume} {61}},\ \bibinfo
  {pages} {012501} (\bibinfo {year} {1999})}\BibitemShut {NoStop}%
\bibitem [{\citenamefont {Theis}\ \emph {et~al.}(2004)\citenamefont {Theis},
  \citenamefont {Thalhammer}, \citenamefont {Winkler}, \citenamefont {Hellwig},
  \citenamefont {Ruff}, \citenamefont {Grimm},\ and\ \citenamefont
  {Denschlag}}]{Theis2004}%
  \BibitemOpen
  \bibfield  {author} {\bibinfo {author} {\bibfnamefont {M.}~\bibnamefont
  {Theis}}, \bibinfo {author} {\bibfnamefont {G.}~\bibnamefont {Thalhammer}},
  \bibinfo {author} {\bibfnamefont {K.}~\bibnamefont {Winkler}}, \bibinfo
  {author} {\bibfnamefont {M.}~\bibnamefont {Hellwig}}, \bibinfo {author}
  {\bibfnamefont {G.}~\bibnamefont {Ruff}}, \bibinfo {author} {\bibfnamefont
  {R.}~\bibnamefont {Grimm}}, \ and\ \bibinfo {author} {\bibfnamefont {J.~H.}\
  \bibnamefont {Denschlag}},\ }\href {\doibase 10.1103/PhysRevLett.93.123001}
  {\bibfield  {journal} {\bibinfo  {journal} {Physical Review Letters}\
  }\textbf {\bibinfo {volume} {93}} (\bibinfo {year} {2004}),\
  10.1103/PhysRevLett.93.123001},\ \Eprint {http://arxiv.org/abs/0404514}
  {arXiv:0404514 [cond-mat]} \BibitemShut {NoStop}%
\bibitem [{\citenamefont {Miller}\ \emph {et~al.}(1993)\citenamefont {Miller},
  \citenamefont {Cline},\ and\ \citenamefont {Heinzen}}]{Miller1993}%
  \BibitemOpen
  \bibfield  {author} {\bibinfo {author} {\bibfnamefont {J.~D.}\ \bibnamefont
  {Miller}}, \bibinfo {author} {\bibfnamefont {R.~A.}\ \bibnamefont {Cline}}, \
  and\ \bibinfo {author} {\bibfnamefont {D.~J.}\ \bibnamefont {Heinzen}},\
  }\href {\doibase 10.1103/PhysRevLett.71.2204} {\bibfield  {journal} {\bibinfo
   {journal} {Physical Review Letters}\ }\textbf {\bibinfo {volume} {71}},\
  \bibinfo {pages} {2204} (\bibinfo {year} {1993})}\BibitemShut {NoStop}%
\bibitem [{\citenamefont {Pellegrini}\ \emph {et~al.}(2008)\citenamefont
  {Pellegrini}, \citenamefont {Gacesa},\ and\ \citenamefont
  {C{\^{o}}t{\'{e}}}}]{Pellegrini2008}%
  \BibitemOpen
  \bibfield  {author} {\bibinfo {author} {\bibfnamefont {P.}~\bibnamefont
  {Pellegrini}}, \bibinfo {author} {\bibfnamefont {M.}~\bibnamefont {Gacesa}},
  \ and\ \bibinfo {author} {\bibfnamefont {R.}~\bibnamefont
  {C{\^{o}}t{\'{e}}}},\ }\href {\doibase 10.1103/PhysRevLett.101.053201}
  {\bibfield  {journal} {\bibinfo  {journal} {Physical Review Letters}\
  }\textbf {\bibinfo {volume} {101}},\ \bibinfo {pages} {053201} (\bibinfo
  {year} {2008})}\BibitemShut {NoStop}%
\bibitem [{\citenamefont {Chen}\ \emph {et~al.}(2014)\citenamefont {Chen},
  \citenamefont {Zhu}, \citenamefont {Li}, \citenamefont {Qian},\ and\
  \citenamefont {Wang}}]{Chen2014}%
  \BibitemOpen
  \bibfield  {author} {\bibinfo {author} {\bibfnamefont {T.}~\bibnamefont
  {Chen}}, \bibinfo {author} {\bibfnamefont {S.}~\bibnamefont {Zhu}}, \bibinfo
  {author} {\bibfnamefont {X.}~\bibnamefont {Li}}, \bibinfo {author}
  {\bibfnamefont {J.}~\bibnamefont {Qian}}, \ and\ \bibinfo {author}
  {\bibfnamefont {Y.}~\bibnamefont {Wang}},\ }\href {\doibase
  10.1103/PhysRevA.89.063402} {\bibfield  {journal} {\bibinfo  {journal}
  {Physical Review A}\ }\textbf {\bibinfo {volume} {89}},\ \bibinfo {pages}
  {063402} (\bibinfo {year} {2014})}\BibitemShut {NoStop}%
\bibitem [{\citenamefont {Dion}\ \emph {et~al.}(2001)\citenamefont {Dion},
  \citenamefont {Drag}, \citenamefont {Dulieu}, \citenamefont {{Laburthe
  Tolra}}, \citenamefont {Masnou-Seeuws},\ and\ \citenamefont
  {Pillet}}]{Dion2001}%
  \BibitemOpen
  \bibfield  {author} {\bibinfo {author} {\bibfnamefont {C.~M.}\ \bibnamefont
  {Dion}}, \bibinfo {author} {\bibfnamefont {C.}~\bibnamefont {Drag}}, \bibinfo
  {author} {\bibfnamefont {O.}~\bibnamefont {Dulieu}}, \bibinfo {author}
  {\bibfnamefont {B.}~\bibnamefont {{Laburthe Tolra}}}, \bibinfo {author}
  {\bibfnamefont {F.}~\bibnamefont {Masnou-Seeuws}}, \ and\ \bibinfo {author}
  {\bibfnamefont {P.}~\bibnamefont {Pillet}},\ }\href {\doibase
  10.1103/PhysRevLett.86.2253} {\bibfield  {journal} {\bibinfo  {journal}
  {Phys. Rev. Lett.}\ }\textbf {\bibinfo {volume} {86}},\ \bibinfo {pages}
  {2253} (\bibinfo {year} {2001})}\BibitemShut {NoStop}%
\bibitem [{\citenamefont {Pechkis}\ \emph {et~al.}(2007)\citenamefont
  {Pechkis}, \citenamefont {Wang}, \citenamefont {Huang}, \citenamefont
  {Eyler}, \citenamefont {Gould}, \citenamefont {Stwalley},\ and\ \citenamefont
  {Koch}}]{Pechkis2007}%
  \BibitemOpen
  \bibfield  {author} {\bibinfo {author} {\bibfnamefont {H.~K.}\ \bibnamefont
  {Pechkis}}, \bibinfo {author} {\bibfnamefont {D.}~\bibnamefont {Wang}},
  \bibinfo {author} {\bibfnamefont {Y.}~\bibnamefont {Huang}}, \bibinfo
  {author} {\bibfnamefont {E.~E.}\ \bibnamefont {Eyler}}, \bibinfo {author}
  {\bibfnamefont {P.~L.}\ \bibnamefont {Gould}}, \bibinfo {author}
  {\bibfnamefont {W.~C.}\ \bibnamefont {Stwalley}}, \ and\ \bibinfo {author}
  {\bibfnamefont {C.~P.}\ \bibnamefont {Koch}},\ }\href {\doibase
  10.1103/PhysRevA.76.022504} {\bibfield  {journal} {\bibinfo  {journal} {Phys.
  Rev. A}\ }\textbf {\bibinfo {volume} {76}},\ \bibinfo {pages} {022504}
  (\bibinfo {year} {2007})}\BibitemShut {NoStop}%
\end{thebibliography}%


%
\end{document}